\documentclass[%
 aip,
 amsmath,amssymb,
 reprint,%
]{revtex4-1}

\usepackage{graphicx}
\usepackage{dcolumn}
\usepackage{bm}
\usepackage{multirow}
\usepackage{subcaption}
\usepackage{mathtools}

\usepackage[utf8]{inputenc}
\usepackage[T1]{fontenc}
\usepackage{mathptmx}
\usepackage{etoolbox}

\graphicspath{ {./Paper_Figures} }

\makeatletter
\def\@email#1#2{%
 \endgroup
 \patchcmd{\titleblock@produce}
  {\frontmatter@RRAPformat}
  {\frontmatter@RRAPformat{\produce@RRAP{*#1\href{mailto:#2}{#2}}}\frontmatter@RRAPformat}
  {}{}
}%
\makeatother
\begin{document}

\preprint{AIP/123-QED}

\title[CCP Argon discharges over a range of frequencies.]{Particle-in-cell simulation of a 50~mTorr capacitively coupled argon discharge over a range of frequencies}
\author{Saurabh Simha}
 \email{ssimha@purdue.edu}
\affiliation{School of Aeronautics and Astronautics, Purdue University, West Lafayette, Indiana - 47906, USA}

\author{Sarveshwar Sharma}%
\affiliation{Institute for Plasma Research and HBNI, Gandhinagar - 382428, India}

\author{Alexander Khrabrov}
\author{Igor Kaganovich}
\affiliation{Princeton Plasma Physics Laboratory, Princeton, New Jersey - 08536, USA}

\author{Jonathan Poggie}
\author{Sergey Macheret}
\affiliation{School of Aeronautics and Astronautics, Purdue University, West Lafayette, Indiana - 47906, USA}

\date{7 May 2023}

\begin{abstract}
The effect of driving frequency in the range of 13.56 MHz to 73 MHz on electron energy distribution and electron heating modes in a 50 mTorr capacitively coupled argon plasma discharge is studied using 1D-3V particle-in-cell simulations. Calculated electron energy probability functions exhibit three distinct ``temperatures'' for low-, mid-, and high-energy electrons. When compared to published experimental data, the calculated probability functions show a reasonable agreement for the energy range resolved in the measurements (about 2 eV to 10 eV). Discrepancies outside this range lead to differences between computational and experimental values of the electron number density determined from the distribution functions, but the predicted effective electron temperature is within 25\% of experimental values. The impedance of the discharge is interpreted in terms of a homogeneous equivalent circuit model and the driving frequency dependence of the inferred combined sheath thickness is found to obey a known, theoretically-derived, power law. The average power transferred from the field to the electrons (electron heating) is computed, and a region of negative heating near the sheath edge, particularly at higher driving frequencies, is identified. Analysis of the electron momentum equation shows that electron inertia, which would average to zero in a linear regime, is responsible for negative values of power deposition near the sheath edge at high driving frequencies due to the highly nonlinear behavior of the discharge.
\end{abstract}

\maketitle

\section{\label{sec:intro}Introduction}

Advanced chip design, for applications in areas such as nano-optics, artificial intelligence, and bio-sensing, presents challenges during the processing steps for fabricating sub-nanometer features on the silicon wafer. In the plasma etching steps, which are used for the removal of materials, this process requires optimization and control over etch rate, uniformity, and surface precision.\cite{lee2014grand} These critical parameters are strongly affected by various factors including plasma generation methods, gas-phase chemistry, and surface conditions. A traditional method of plasma generation involves applying radio-frequency voltages, typically at 13.56 MHz, between sets of parallel electrodes, in a configuration known as a capacitively coupled plasma (CCP) discharge.\cite{lieberman2005principles,grill1994cold,raizer1995rfcd} 

Plasma discharges operated in the very high frequency (VHF) range (30~MHz to 300~MHz), however, have been shown to offer numerous benefits. VHF discharges result in a high ion current, increasing the ion yield to the substrate; and a reduced sheath thickness, aiding in ion directionality\cite{surendra1991capacitively, colgan1994very, vahedi1993verification} and enabling micro- and nano-scale etching with a high aspect ratio.\cite{ohtake2003highly,misra1998plasma} A high ion yield allows the operation of the discharge at a lower pressure and a lower voltage amplitude, reducing the required power input and lessening damage to the substrate.

Due to high-frequency operation, the shape of the Ion Energy Distribution Function (IEDF) impinging on the wafer surface is narrow. On the other hand, electrons which are strongly heated by the sheath oscillations penetrate into bulk plasma, and drive the gas-phase plasma chemistry via collisions with the background gas. These collisions lead to a modification in the Electron Energy Distribution Function (EEDF). Published experiments and simulations have confirmed that low pressure discharges exhibit a bi-Maxwellian electron energy distribution, where a major population of electrons has low mean energy, but a small population of electrons has very high mean energy.\cite{godyak1990abnormally,colgan1994very,abdel2003electron,abdel2003influence,abdel2013combined,sharma2016effect,sharma2019influence,sharma2020driving,sharma2019electric,sharma2018influence} These fast electrons assist in ionization and sustain the discharge. Since the overall population of electrons has low mean energy, the plasma can be operated with a relatively low input power.

The nature of electron energy distribution functions (EEDF) is determined by the mechanism through which electrons gain and lose energy. Typically, there are three modes\cite{abdel2003influence} of electron heating: collisional or ohmic heating, collisionless heating near the sheath, and secondary electron-induced heating. The third mechanism is dominant in high voltage discharges.\cite{wen2019secondary} At low driving voltages, the dominant heating mechanism varies between collisionless heating at lower pressures and ohmic heating at higher pressures. Two interpretations of collisionless heating are known. Collisionless heating can be the result of energy transfer between the oscillating sheath edge and the electrons near the sheath edge, also called stochastic heating;\cite{raizer1995rfcd} or due to compression of the electron cloud near the sheath edge, also called pressure heating.\cite{turner1995pressure} Electrons lose their energy through inelastic collisions with the neutral gas particles and absorption at the boundaries.

CCP discharges are nonlinear due to the space-time dependence of the driving electric field and the nonuniform ion density in the sheath, which makes theoretical analysis challenging. Researchers have relied on experimental and computational techniques to study discharge behavior. Godyak and Piejak\cite{godyak1990abnormally} studied the transition of the electron energy distribution from a two-temperature distribution at lower pressures to a Druyvesteyn-like distribution at higher pressures (above 1~Torr) when the driving frequency was 13.56~MHz. They concluded that this transition was caused by the change in electron heating mode from collisionless heating at lower pressures to ohmic heating at higher pressures.

The need to reduce the power required to operate the discharge and increase the yield of ions has prompted the study of the frequency dependence of the behavior of these discharges. Abdel-Fattah and Sugai\cite{abdel2003electron,abdel2003influence} conducted experiments at 50~mTorr to 150~mTorr at various driving frequencies for helium and argon CCP discharges, and observed a transition from a Druyvesteyn-type distribution at low frequencies to a bi-Maxwellian distribution at high frequencies. The transition was associated with an increase in electron number density at the center of the discharge and a decrease in effective electron temperature. As the frequency was increased, the temperature of the bulk electrons in the center of the discharge decreased, whereas the temperature of the tail electrons increased.

In this paper, we present results of Particle-In-Cell~(PIC) simulations for the conditions studied experimentally by Abdel-Fattah and Sugai.\cite{abdel2003electron,abdel2003influence} They determined the electron energy distribution for a wide range of frequencies, a parametric effect has not been widely studied. Colgan et. al.\cite{colgan1994very} performed experiments on an argon discharge at 250~mTorr for various frequencies, but the electron energy distributions were not studied. Godyak et al.\cite{godyak1990abnormally} studied electron energy distributions for high pressures up to 3~Torr, but only at 13.56~MHz driving frequency.

Nonlinearities in capacitively coupled radio frequency~(CCRF) discharges appear in the form of higher harmonic oscillations of the electric field due to production of energetic electron beams.\cite{wilczek2015effect,wilczek2018disparity,sharma2019electric,sharma2020electric,sharma2020driving,sharma2018plasma} In low pressure discharges, interaction of the electron beams with moving sheaths leads to collisionless heating. Above a critical driving frequency, this process can lead to a drastic increase in the electron density.\cite{wilczek2015effect,sharma2018plasma} Power spectra of the electric field at various locations in the discharge show the presence of higher harmonic oscillations, which aid in the confinement and enhance the heating of electrons.\cite{sharma2020electric,sharma2020driving} At higher pressures, Wilczek et al.\cite{wilczek2015effect} found that there was no drastic increase in the electron density and that the variation of the electron density with driving frequency was closer to the theoretical estimate\cite{lieberman2005principles,chabert_braithwaite_2011} of $n_{\rm e} \propto f^2$.

Because of computational cost, many PIC simulations\cite{sharma2016effect,sharma2019influence,sharma2020driving,sharma2019electric,turner2013simulation,wilczek2015effect,wilczek2018disparity,sharma2020electric} have been conducted at lower pressures (below 10~mTorr). Since higher pressure results in high ion yield for materials processing applications, the current study focuses on conducting simulations for a 50~mTorr argon plasma at various frequencies and comparing the results with published experimental data.\cite{abdel2003electron,abdel2003influence} In addition to the frequency dependance of the discharge behavior, we also investigate the influence of nonlinearity on electron heating and electron density.

The paper is organized as follows. In section 2, we discuss the numerical procedures used to solve the governing equations (Boltzmann equation for the species and Poisson equation for the electric potential) along with the simulation parameters used for this study. In the first part of section 3, we discuss the electron energy probability functions obtained from the simulations at various driving frequencies. We calculate the electron densities and the effective electron temperatures, and compare the distributions and the aforementioned calculated plasma properties to results from published experimental data. In the second part of section 3, we use a simple homogeneous lumped-parameter model to determine the sheath thickness. In the last part of section 3, the power deposition by the field to the electrons is analyzed and all the features observed are discussed in detail. Finally, a summary of the results are provided. 

\section{\label{sec2:procedure}Procedure}
The simulations are carried out using a modified version of the EDIPIC (electrostatic direct-implicit particle-in-cell) code\cite{sydorenko2006particle} developed at the Princeton Plasma Physics Laboratory. The code simulates one dimension in physical space and three dimensions in velocity space (1D-3V). This code has been benchmarked for various cases and shown to be accurate in predicting the properties of a discharge over a wide range of pressures.\cite{sharma2018plasma,sheehan2013kinetic,patil2022electron,sharma2018spatial,sharma2022investigating}

\begin{figure}[htb]
    \centering
    \includegraphics[width=0.4\textwidth]{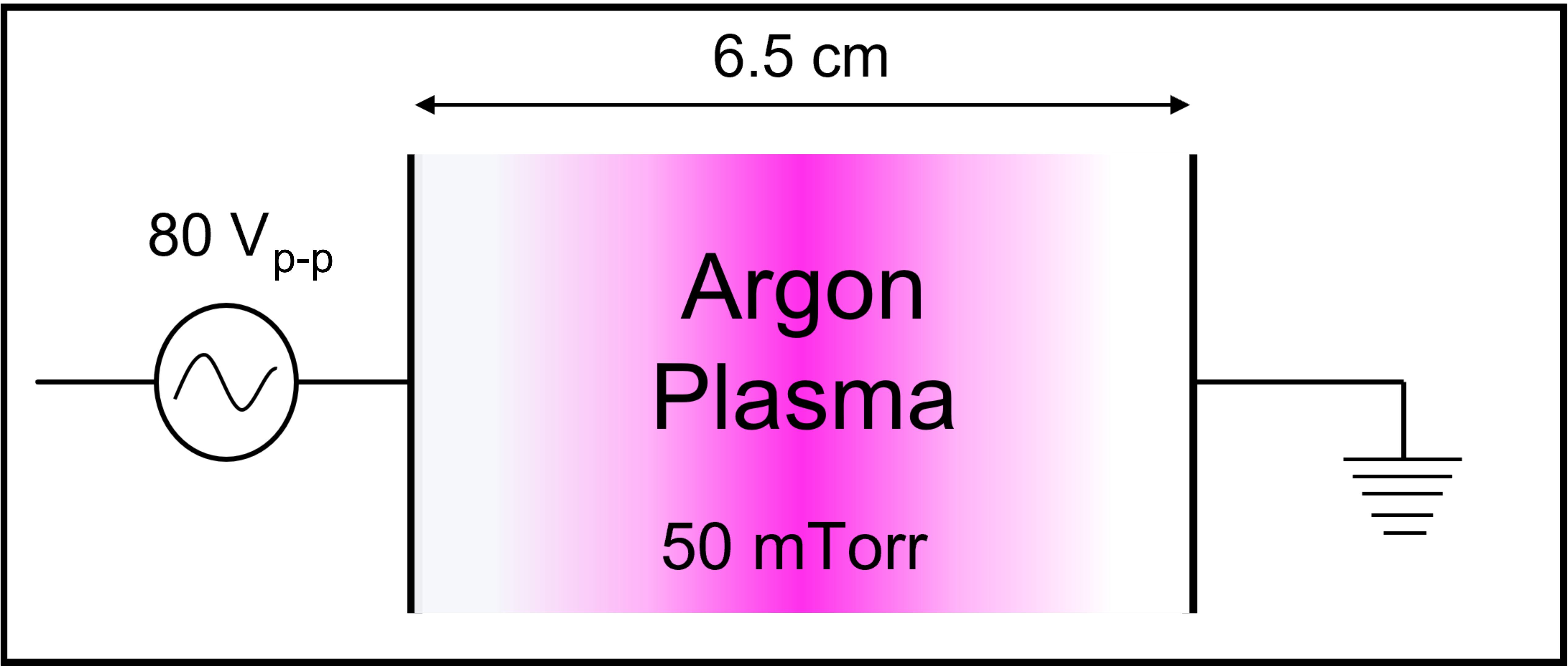}
    \caption{Schematic diagram of the simulation domain.}
    \label{fig1:domain}
\end{figure}

In order to carry out a parametric study of the influence of driving frequency,
the conditions for the simulations in this study were chosen to match those of Abdel-Fattah and Sugai's published experiments.\cite{abdel2003influence} The simulation parameters are summarized in table~\ref{tab1:simpar} and a schematic diagram of the computational domain is provided in figure~\ref{fig1:domain}. An argon plasma is operated using an 80~V (peak-to-peak) sinusoidal voltage ($V_{RF}=40 \sin \left(2 \pi f t \right)$) applied to the left electrode at  driving frequencies between 13.56~MHz and 73~MHz. The right electrode is grounded. The gap between the electrodes is 6.5~cm and the neutral gas pressure is 50~mTorr (about 6.7~Pa). The associated neutral gas density is $1.61 \times 10^{21}\ {\rm m^{-3}}$ which is uniformly distributed throughout the simulation domain. The gas temperature is 300~K. Since the maximum magnitude of potential difference applied between the electrodes is 40~V, the boundary conditions for the walls are set to be 100~\% absorbing. Hence, the values of secondary emission coefficient and the effective reflection coefficient are taken to be 0.

\begin{table}
\caption{\label{tab1:simpar}Simulation parameters.}
\footnotesize
\begin{ruledtabular}
\begin{tabular}{ll}
Parameter   &   Values\\
\hline
Driving Frequencies (MHz)                 &   13.56, 27, 37, 44, 50, 57, 65, 73\\
Cell Size (${\rm \mu m}$)                 &   $16.62\; \left(\Delta x << \lambda_{\rm D} \approx 300\;{\rm \mu m} \right)$\\
Time Step (ps)                            &   $1 \;\left(\Delta t << 1/\omega_{\rm p} \approx 100\;{\rm ps} \right)$\\
Number of macroparticles per cell         &   200\\
\multirow{4}{10em}{Particle Interactions} &   \textbf{electron-neutral}: elastic, excitation,\\
                                          & and ionizing collisions.\\
                                          &   \textbf{ion-neutral}: elastic and charge\\
                                          & exchange collisions.\\
\end{tabular}
\end{ruledtabular}
\end{table}

\begin{figure*}[t]
    \centering
    \includegraphics[width=0.9\textwidth]{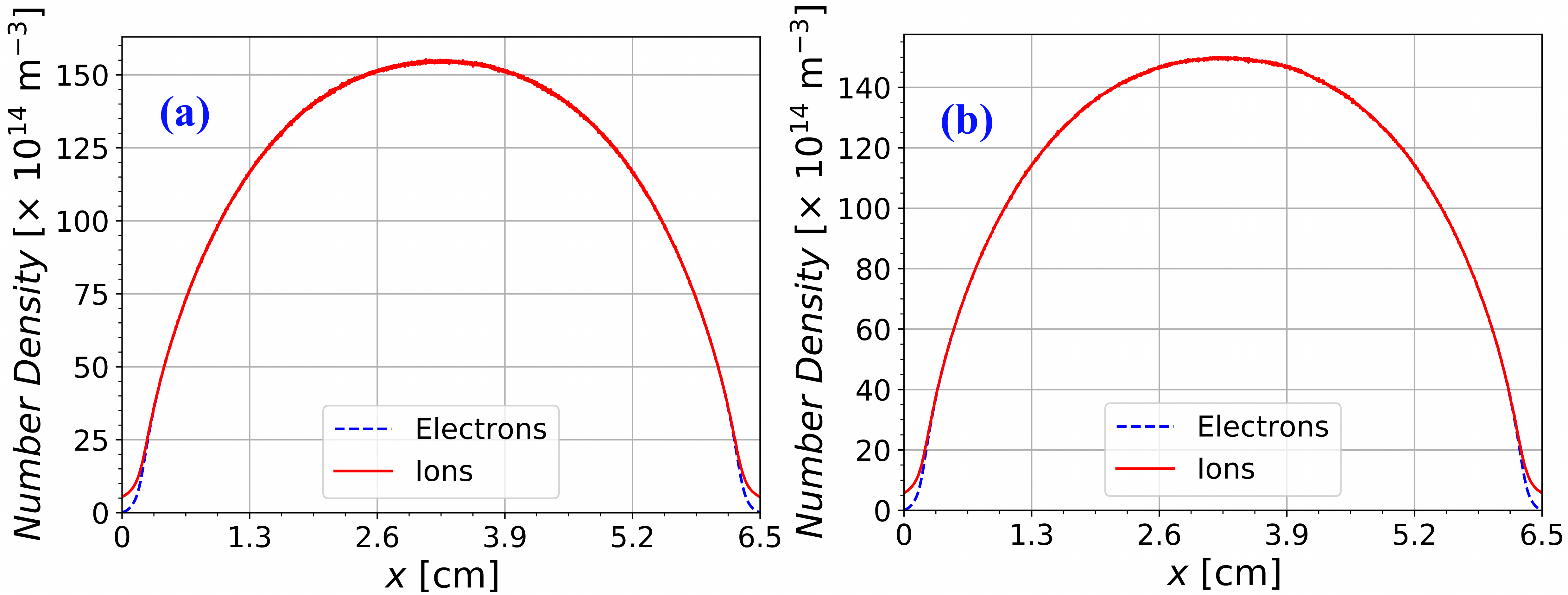}
    \caption{Profiles of time average electron and ion densities for 73~MHz when the grid size is (a) 16.62 ${\rm \mu m}$ and (b) 41.5 ${\rm \mu m}$. }
    \label{fig2:den_grid}
\end{figure*}
The code employs direct implicit or explicit integration for the particle Boltzmann equations and solves the Poisson equation for electric potential at intermediate steps. For all the simulations in this work, we have used the explicit scheme. Message Passing Interface (MPI) library calls enable parallel computations, and the code writes various raw data and integrated quantities at regular intervals. It includes models for electron-neutral elastic, excitation, and ionizing collisions, as well as ion-neutral elastic and charge exchange collisions.
\begin{figure*}
    \centering
    \includegraphics[width=0.9\textwidth]{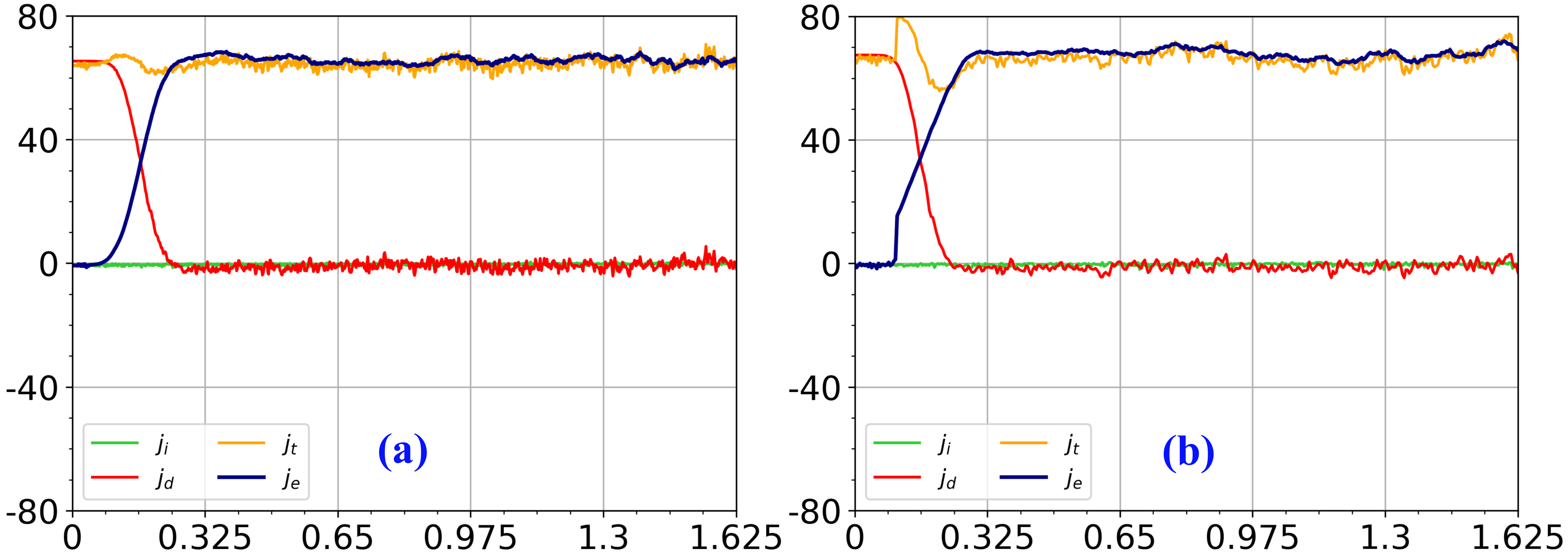}
    \caption{Profiles of current at the beginning of the voltage cycle ($\omega t = 0$) for 73 ~MHz when the grid size is (a) 16.62 ${\rm \mu m}$ and (b) 41.5 ${\rm \mu m}$. }
    \label{fig3:curr_grid}
\end{figure*}
The choice of using an electrostatic, one-dimensional  model is made due to the dimensions of the configuration studied here. Even for the highest driving frequency examined, standing wave effects and skin effects are minimal, establishing plasma uniformity and symmetry in the radial direction.\cite{Lieberman_2002,raizer1995rfcd} The limits on the discharge dimensions for minimal electromagnetic effects and skin effects are set by the characteristic free space wavelength, $\lambda_0$, and the penetration depth of an oscillating field into the plasma, $\delta^{'}$.
\begin{eqnarray}
\lambda_0 = \frac{2 \pi c}{\omega} \quad {\rm and} \quad
\delta^{'} = \frac{c}{f \sqrt{|\Re\left(\epsilon_{\rm r}\right)|}}.
\label{eq1:char_param}
\end{eqnarray}
where c is the speed of light in vacuum, $\omega = 2 \pi f$ is the driving frequency in rad/s, and $\Re\left(\epsilon_{\rm r}\right)$ denotes the real part of the plasma permittivity (defined later in section~\ref{sec:j_e.E}). When calculated for the highest driving frequency in this study, 73 MHz, $\lambda_0 = 411\ {\rm cm}$ and $\delta^{'} = 49.19\ {\rm cm}$, which are greater than the dimensions of the discharge, namely, the gap between the electrodes (6.5 cm) and the radius of the electrodes (7.5 cm). 

For the calculations, the number of macroparticles per cell is chosen to be 200. The gap between the electrodes is divided into 3911 equal sized cells, $16.62\ {\rm \mu m}$ wide. The cell size is chosen to satisfy the limit imposed by the Debye length ($\lambda_{\rm D} \approx 300\ {\rm \mu m}$). 
\begin{figure*}[ht]
    \centering
    \includegraphics[width=0.9\textwidth]{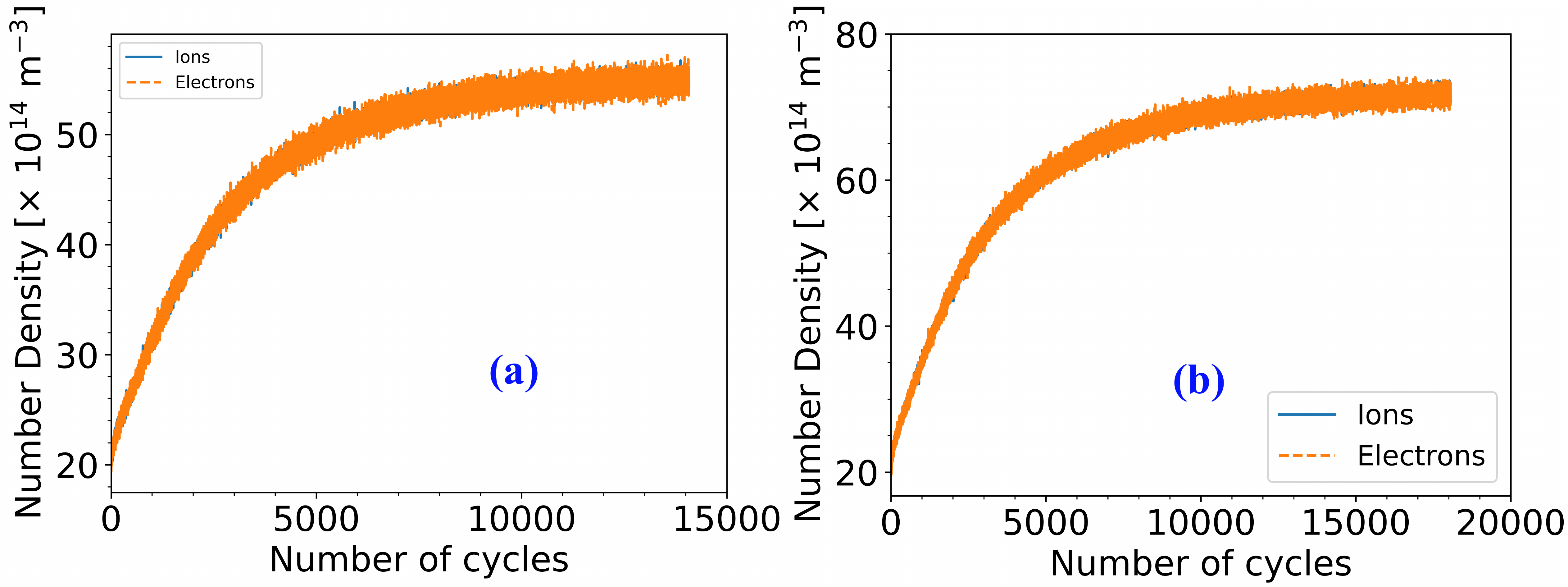}
    \caption{Variation of electron and ion densities at the center of the discharge gap with number of RF cycles elapsed for (a) 44 MHz, and (b) 50 MHz.}
    \label{fig4:convergence}
\end{figure*}
An additional study with 1566 cells in the domain, each of width $41.5\ {\rm \mu m}$, was done to study the effect of grid resolution. Figure \ref{fig2:den_grid} shows the time average number density profiles for the 73 MHz driving frequency case for the two different grid sizes and it can be seen that the finer grid resolves the profile of density in the quasineutral region, whereas the coarse grid results in a lower density in the center.

Figure \ref{fig3:curr_grid} shows the profiles of electron current at the beginning of the input voltage cycle ($\omega t = 0$ according to the defined voltage) for the two grid sizes. The coarser grid produces spurious oscillations in the current near the sheath edge (below 0.325 cm) and in the bulk (about 0.975 cm to 1.625 cm) because of the lack of spatial resolution. A cell size of $16.62\ {\rm \mu m}$ was used for the remaining results presented in this article.

The calculations were carried out on computing resources provided by the Purdue University Community Cluster Program.\cite{McCartney2014} Initially, a case was run for thousands of RF cycles to allow it to reach a periodic state. Convergence was confirmed by monitoring the time variation of  electron and ion densities at the center of the domain after the initial run. In a periodic state, the value of electron and ion densities at the center of the domain should be constant, as shown in figure \ref{fig4:convergence}. After a periodic state was achieved, the simulation was run for an additional 100~RF cycles to obtain statistical data. The time steps were chosen to be smaller than period of plasma oscillations divided by $2\pi$ ($1/\omega_{\rm p} \approx 100\ {\rm ps})$.

\section{Results}
\subsection{\label{sec:distfunc}Distribution Functions}

\begin{figure*}[htb]
    \centering
    \includegraphics[width=0.9\textwidth]{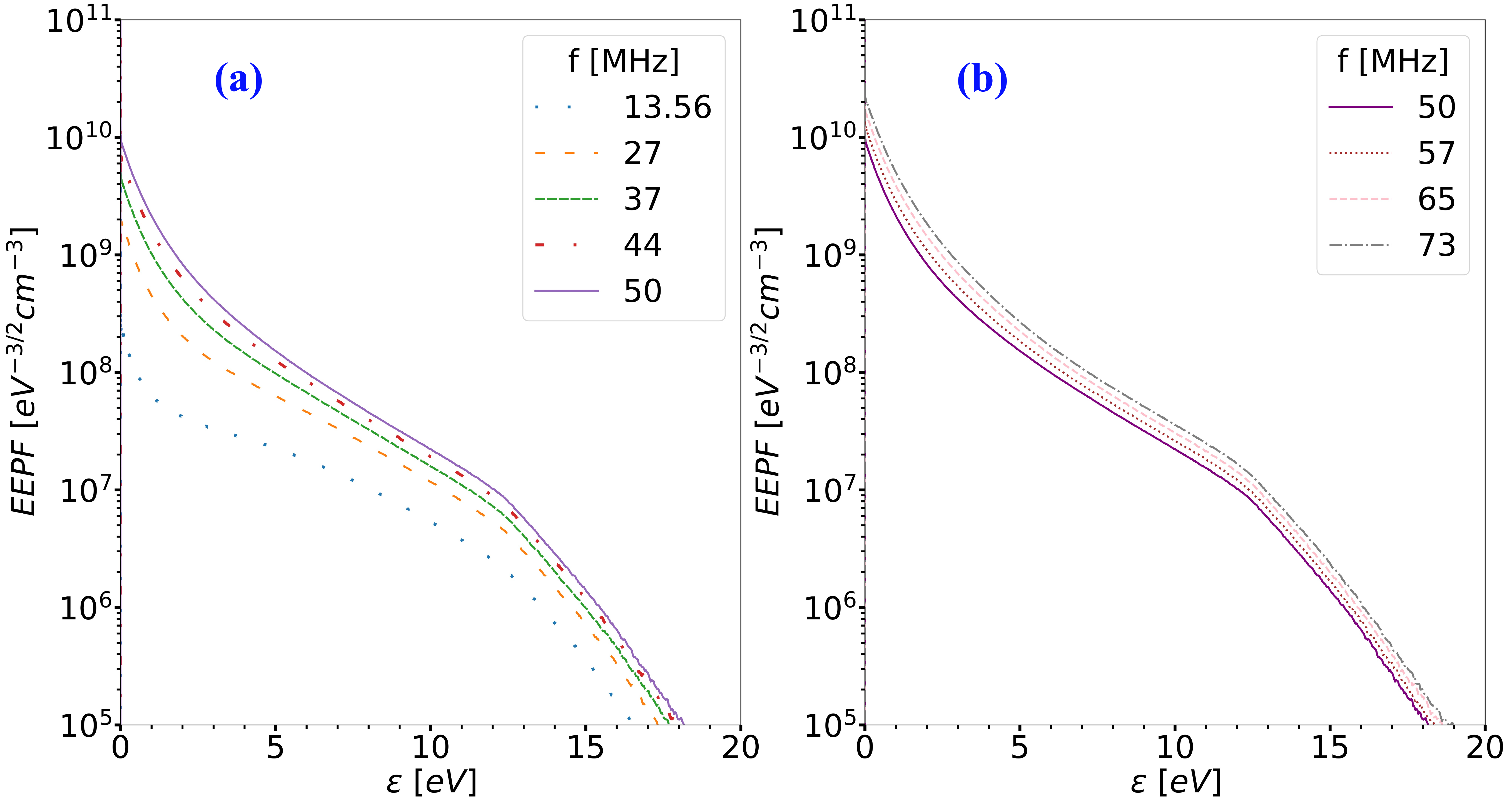}
    \caption{Electron energy probability functions at the center of the discharge gap for (a) 13.56~MHz to 50~MHz, and (b) 50~MHz to 73~MHz.}
    \label{fig5:distvsfreq}
\end{figure*}

Figure~\ref{fig5:distvsfreq} shows the calculated electron energy probability functions (EEPF) at the center of the discharge gap for each case. (See table~\ref{tab1:simpar} for the list of driving frequencies.) The data are normalized such that the EEPF, when multiplied by the square root of the energy and integrated over energy, gives the electron number density. Plotted in this form, the shape of the EEPF for each case exhibits three regions of roughly constant slope, one at very low electron energies, one in the mid-energy range, and finally one at high electron energies. Non-Maxwellian energy distributions in capacitively coupled discharges form due to nonlocal kinetic effects, a phenomenon which has been studied previously for low background gas pressure.\cite{kaganovich1992low,kaganovich1992space,berezhnoi1998fast,berezhnoi1998generation}

As frequency increases, the shape of the distribution evolves, and the curve shifts upwards because of an increase in electron number density. For the cases from 13.56~MHz to 44~MHz, figure~\ref{fig5:distvsfreq}(a), the population of electrons in the low-energy region increases much faster than the population in the higher energy range. This behavior is because of the production of low-energy electrons in ionization processes. For the cases from 50~MHz to 73~MHz, figure~\ref{fig5:distvsfreq}(b), the population of both the bulk and the tail portions of the EEPF increases proportionally.

\begin{figure*}[htb]
  \centering
  \includegraphics[width=0.9\textwidth]{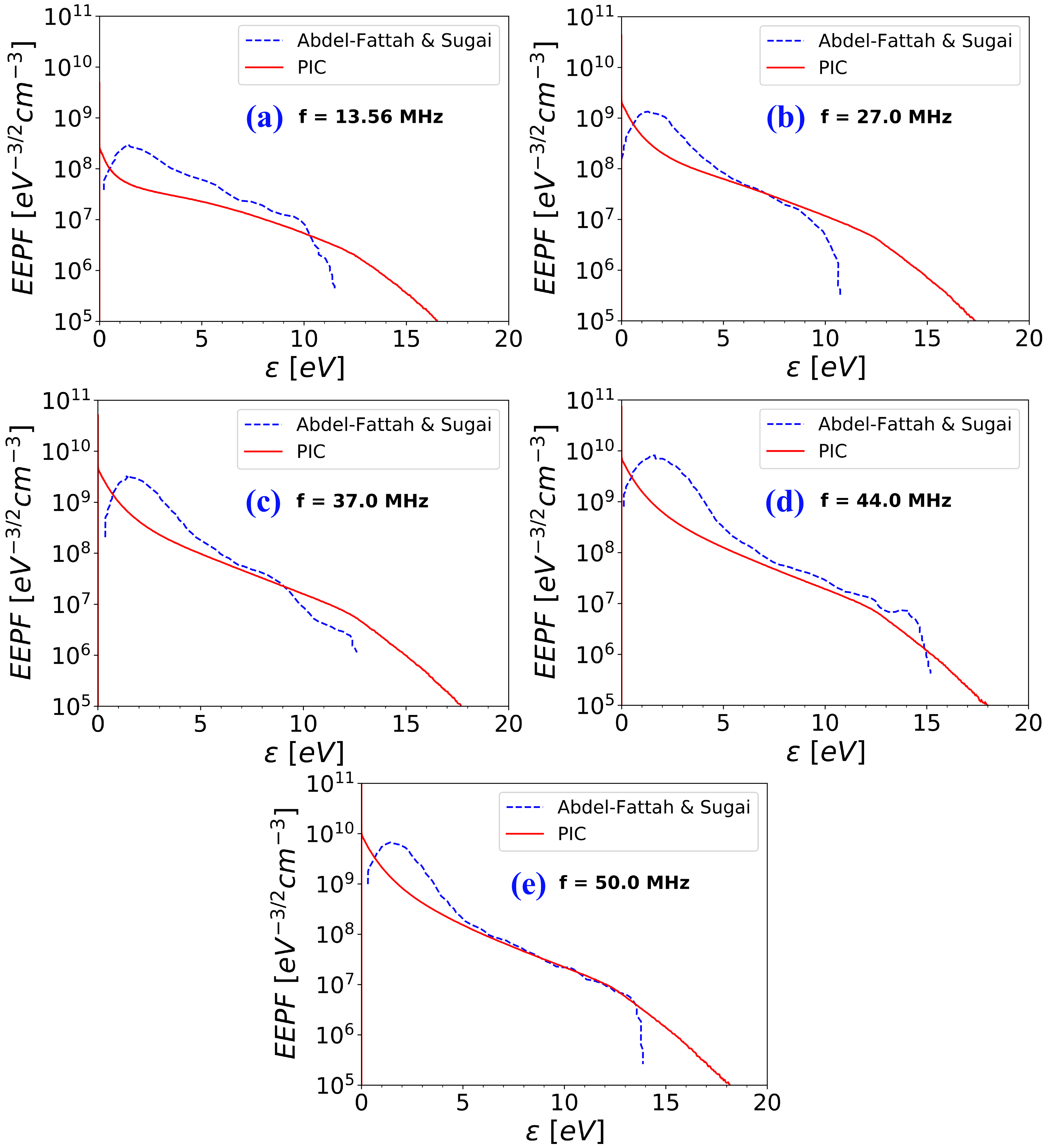}
  \caption{Comparison of predicted and measured\cite{abdel2003influence} electron energy probability functions for driving frequencies in the range of 13.56 MHz to 50 MHz.}
  \label{fig6:distcomp}
\end{figure*}

\begin{figure*}[htb]
    \centering
    \includegraphics[width=0.9\textwidth]{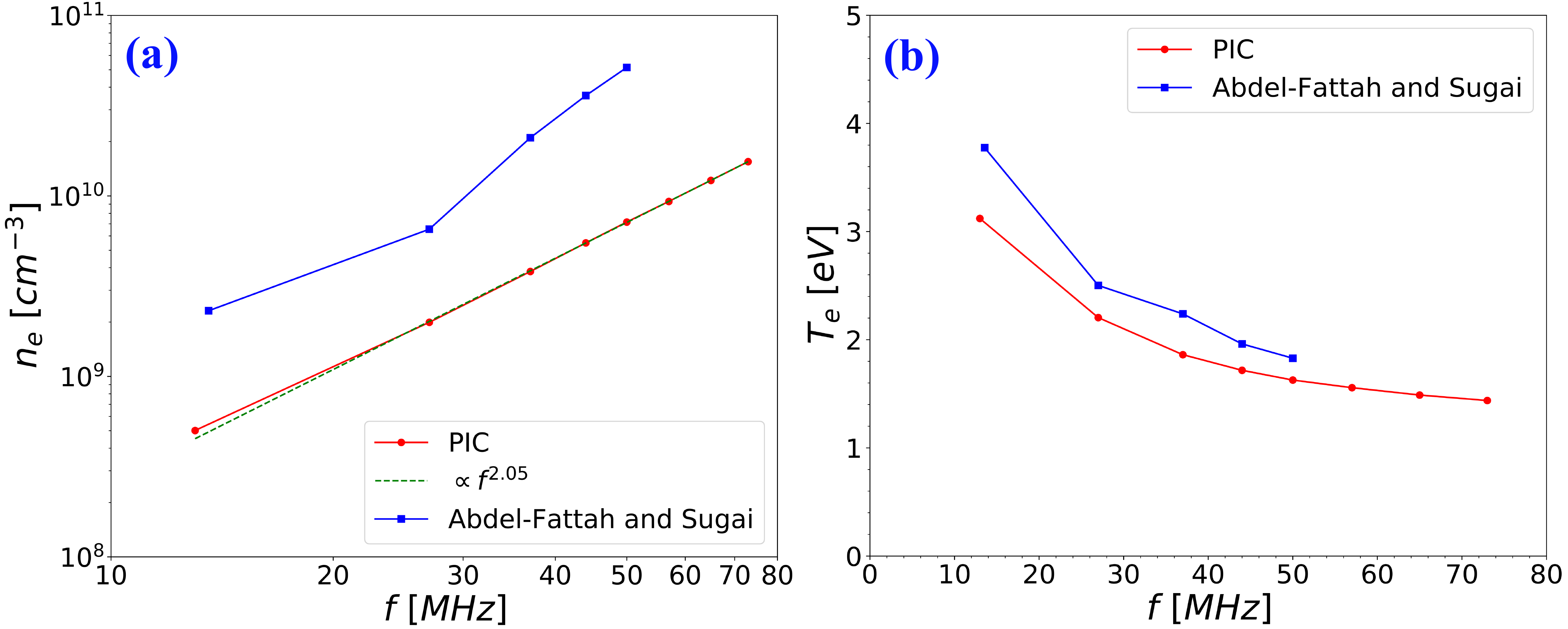}
    \caption{Electron number density (a) and effective electron temperature (b) as determined from the EEPF at the center of the discharge. Results of the present simulations are compared to those of published experiments.\cite{abdel2003influence}}
    \label{fig7:ndte}
\end{figure*}

Figure~\ref{fig6:distcomp} compares the distribution functions obtained from the present simulations to the corresponding experimental data.\cite{abdel2003influence} For the purpose of comparison of simulation results with the experimental results, the data in figure~\ref{fig6:distcomp} and figure~\ref{fig7:ndte} are plotted in the units of $\rm cm^{-3}$.

In comparing the predictions to the measurements, it must be kept in mind that the probes used in the experiments only provide accurate data for a limited electron energy range.\cite{abdel2003electron} The experimental data are not accurate for very low and very high energies. Furthermore, the experimental data in Reference~\cite{abdel2003influence} were extrapolated to zero for the low energy values, which results in an unphysical local maximum at low energy for the experimental distribution. As a result, at low driving frequencies, the experimental distributions do not exhibit the first constant slope region at low electron energies which are observed in the simulated distribution functions. At high electron energies (greater than about 12 eV to 15 eV), the experimental distributions fall steeply, whereas the computational results show a third region of constant slope.

These experimental limitations also affect the normalization of the distribution, introducing a shift in the experimental EEPF and uncertainty in the electron number density derived from its integral. Nonetheless, the computed values for the EEPF agree reasonably well with experimental measurements in the range of 2~eV to 10~eV for all cases.

Macroscopic quantities such as the electron number density $n_{\rm e}$, and the effective electron temperature $T_{\rm e}$, were calculated at the center of the discharge from the computed EEPF. The electron density was determined from the EEPF as:
\begin{equation}
  \label{eq2:ne}
   n_{\rm e} = \int^\infty_0 \sqrt{\epsilon}\; f_{\rm p} \left( \epsilon \right) \, {\rm d} \epsilon
\end{equation}
where $\epsilon$ denotes electron energy and $f_{\rm p}\left(\epsilon\right)$ is the EEPF. The effective electron temperature (in electron volts) corresponds to the mean electron energy $\left< \epsilon \right>$, and is calculated from the EEPF as:
\begin{equation}
  \label{eq3:te}
  T_{\rm e} = \frac{2 \left< \epsilon \right>}{3} = \frac{2}{3 n_{\rm e}} \; \int^\infty_0 \epsilon^{3/2} \; f_{\rm p} \left( \epsilon \right) \, {\rm d} \epsilon
\end{equation}

Figure~\ref{fig7:ndte} shows the comparison of the electron density and effective electron temperature, calculated from the EEPF using equations~(\ref{eq2:ne}) and (\ref{eq3:te}), between simulation and experiment.\cite{abdel2003influence} The electron number density, figure~\ref{fig7:ndte}(a), increases with frequency, and the predicted rate of increase roughly matches that of the experiments. The magnitude of the predicted electron number density, however, is significantly lower than the reported experimental values. As mentioned earlier, limitations in the experimental energy resolution introduce a shift in the experimental EEPF and uncertainty in the electron number density derived from its integral. From figure~\ref{fig6:distcomp}, we see that there is an additional area under the EEPF curve in the experimental data which does not exist in the simulation results. This uncertainty may, in part, explain the level of discrepancy. 

In the calculations reported in this paper, both secondary emission and electron reflection at the electrodes were disregarded due to the relatively low pressure and voltage. Surface effects at the electrodes were found by Schulenberg et. al.\cite{schulenberg2021multi} to play an important role in contributing to the electron density at relatively high applied voltages (150~V to 350~V). Those authors used a trial and error method to find that a reflection coefficient of 0.7, and a secondary emission coefficient of 0.07, provided a best match for their calculated electron densities to their measured values. To ensure that the discrepancy in electron density between the present simulations and Reference~\cite{abdel2003electron} was not due to surface effects, a similar test for our low voltage discharge cases was conducted. It was found that the electron density increased by only 26\%, insufficient to explain the difference between the simulation and experimental results.

According to a simple theoretical analysis,\cite{lieberman2005principles,chabert_braithwaite_2011} which ignores energy gained by electrons from collisionless heating, electron density is predicted to increase as $n_{\rm e} \propto f^2$. The present computations predict that the electron density increases as $n_{\rm e} \propto f^{2.05}$, similar to the theoretical trend. For the present case, the gas pressure is not low enough for collisionless heating to be a dominant mode of power deposition. The trend of experimentally calculated electron density versus the driving frequency shows a change in the slope of the curve after a driving frequency of 27 MHz; this behavior is absent in the computational curve.

It is to be noted that Wilczek et. al.\cite{wilczek2015effect} studied the effect of background gas pressure on the dependence of bulk electron density versus the driving frequency and found that at pressures above 2~Pa (about 15~mTorr) the drastic increase in electron density after certain driving frequency disappeared, and the trend became closer to the theoretical approximation of $n_{\rm e} \sim f^2$, for which non-local kinetic effects are ignored.\cite{lieberman2005principles,chabert_braithwaite_2011}

The corresponding effective electron temperature data are shown in figure~\ref{fig7:ndte}(b). Here, the predicted values through simulation match the experimental data fairly well. To understand the trends in this figure, we consider the slope of the EEPF. For a Maxwellian distribution, the EEPF and the derivative of its logarithm are:
\begin{eqnarray}
  \label{eq4:eepf}
  f_{\rm p} \left( \epsilon \right) = \frac{\mathcal{C}}{T_{\rm e}^{3/2}} \; \exp \left( -\frac{\epsilon}{T_{\rm e}} \right) \\
  \label{eq5:teslope}
  \frac{{\rm d}}{{\rm d} \epsilon} \log f_{\rm p} \left( \epsilon \right) = -\frac{1}{T_{\rm e}}
\end{eqnarray}
where $\cal{C}$ is the normalization constant. From equation~(\ref{eq5:teslope}), the electron temperature of a Maxwellian distribution is described by the slope of the EEPF on a log scale. If we interpret the regions of nearly constant slope in figure~\ref{fig5:distvsfreq}(a) as quasi-Maxwellian, we see that the effective electron temperature over a broad range of electron energies (about 2~eV to 12~eV) decreases with driving frequency (increasingly negative slope) over the range 13.56~MHz to 50~MHz. At higher frequencies, figure~\ref{fig5:distvsfreq}(b) shows that there is a negligible difference in the slopes of the distribution functions. Hence, above 50~MHz, the decrease in effective electron temperature with increasing driving frequency is small as seen in figure~\ref{fig7:ndte}(b).

\begin{figure*}[htb]
    \centering
    \includegraphics[width=0.9\textwidth]{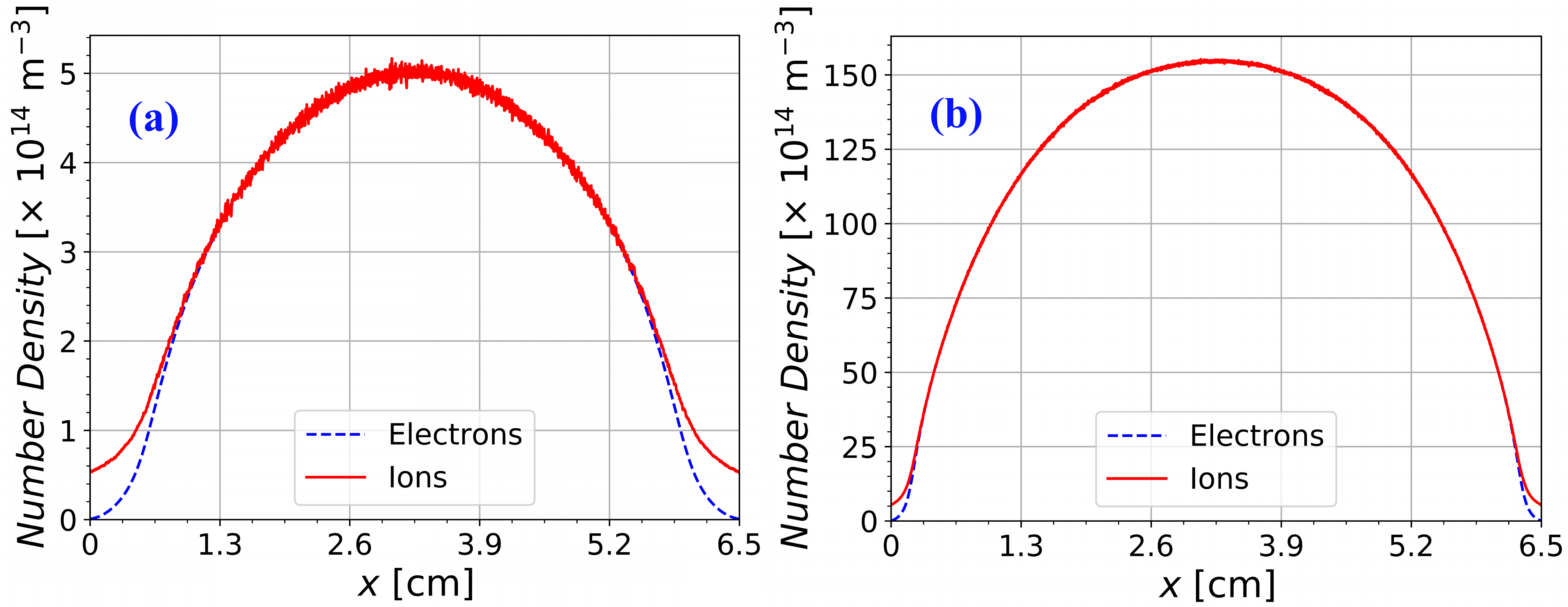}
    \caption{Average electron and ion density profiles in the discharge domain for (a) 13.56 MHz and (b) 73 MHz.}
    \label{fig8:denpro}
\end{figure*}

Figure~\ref{fig8:denpro} shows the profiles of electron and ion densities for the lowest (13.56 MHz) and the highest (73 MHz) driving frequency cases. As the driving frequency increases from 13.56 MHz to 73 MHz, the density at the center also increases from $5 \times 10^{14}\ {\rm m^{-3}}$ to $1.5 \times 10^{16}\ {\rm m^{-3}}$ respectively, as was previously observed, $5 \times 10^{8}\ {\rm cm^{-3}}$ to $1.5 \times 10^{10}\ {\rm cm^{-3}}$ in figure~\ref{fig7:ndte}(a). Ionization increases at higher driving frequencies and causes the plasma to fill up the discharge gap. Consequently, the thickness of the combined sheath also decreases from 10 mm to 3.3 mm as the driving frequency increases from 13.56 MHz to 73 MHz. The thickness of the sheath for all the driving frequencies are calculated from the simulation parameters a simple homogeneous model in the next section.

\subsection{\label{sec:sheath}Equivalent Circuit Model}

In the VHF range, CCP discharges can be modeled by a simple lumped-parameter circuit,\cite{raizer1995rfcd} since the wavelength of the input voltage signal is much larger than all geometric scales involved in the discharge. In this homogeneous model, a uniform quasineutral plasma is sandwiched between two sheaths completely devoid of electrons. The ions are assumed to be stationary, with constant density throughout the discharge gap. In contrast, the electrons are assumed to respond instantaneously to the electric field, and to oscillate at the driving frequency between the electrodes.

\begin{figure}[htb]
    \centering
    \includegraphics[width=0.45\textwidth]{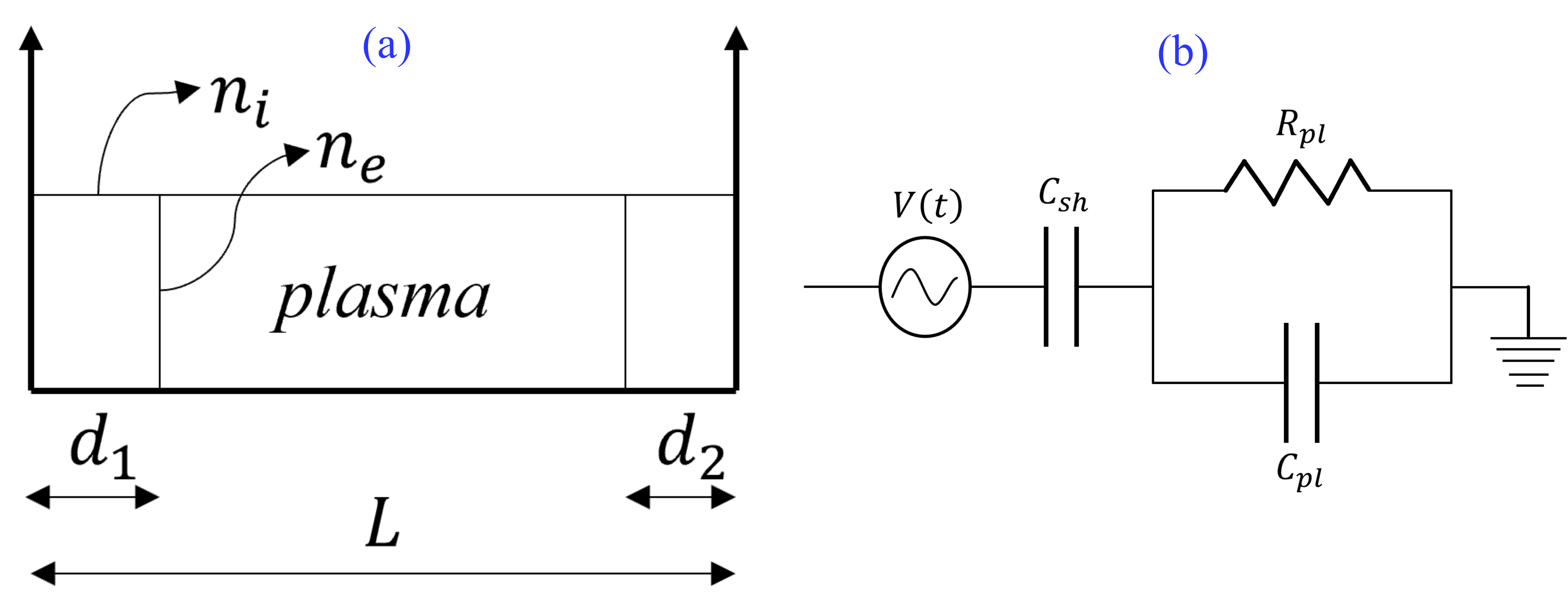}
    \caption{(a) Depiction of homogeneous plasma with uniform ion density and electron-free sheaths. (b) Equivalent circuit with lumped capacitance for the two sheaths and lumped resistance and capacitance for the plasma.}
    \label{fig9:circ}
\end{figure}

The two sheaths are represented by variable gaps of width $d_{\rm 1}$ and $d_{\rm 2}$, as shown in figure~\ref{fig9:circ}(a). The plasma column of thickness $L - \left(d_{\rm 1} + d_{\rm 2}\right)$, where $L$ is the gap between the electrodes, oscillates with the RF frequency, just touching the electrodes.

The equivalent circuit diagram for the discharge gap is shown in figure~\ref{fig9:circ}(b). The two sheaths are combined into a single equivalent capacitance $C_{\rm sh}$ with constant total size $d_{\rm sh}=d_{\rm 1} + d_{\rm 2}$. The plasma column is modeled by a combination of a resistance $R_{\rm pl}$ and a capacitance $C_{\rm pl}$ connected in parallel. A detailed explanation of this model is given in Reference~\cite{raizer1995rfcd}.

The values of the electrical elements in the equivalent circuit can be calculated from the electron number density, driving frequency, and the spatial sizes as:
\begin{eqnarray}
R_{\rm pl} = \frac{\left(\omega^2 + \nu_{\rm m}^2 \right) \left(L - d_{\rm sh} \right)}{\epsilon_{\rm 0} \nu_{\rm m} \omega_{\rm p}^2 S}. \nonumber \\
C_{\rm pl} = \frac{\epsilon_{\rm 0} S}{L-d_{\rm sh}} \left( 1 - \frac{\omega_{\rm p}^2}{\omega^2 + \nu_{\rm m}^2} \right) \quad {\rm and}\quad X_{\rm pl} = \frac{1}{j \omega C_{\rm pl}}.  \label{eq6:circ} \\
C_{\rm sh} = \frac{\epsilon_{\rm 0} \epsilon_{\rm Ar} S}{d_{\rm sh}} \quad {\rm and} \quad X_{\rm sh} = \frac{1}{j \omega C_{\rm sh}}. \nonumber
\end{eqnarray}
where $X$ denotes the reactance of the capacitors, $S$ denotes the electrode area, $\omega = 2 \pi f$ is the driving frequency, $\nu_{\rm m}$ is the electron-neutral collision frequency, $\omega_{\rm p}^2 = e^2 n_{\rm e}/\epsilon_{\rm 0} m_{\rm e}$ is the electron plasma frequency, and $\epsilon_{\rm 0}$ and $\epsilon_{\rm Ar} \approx 1$ denote the vacuum permittivity and the relative permittivity of argon respectively.  The overall impedance of the plasma is:
\begin{equation}
Z = X_{\rm sh} + \frac{R_{\rm pl}X_{\rm pl}}{R_{\rm pl} + X_{\rm pl}}.
\label{eq7:imped}
\end{equation}
\begin{figure*}[htb]
    \centering
    \includegraphics[width=0.9\textwidth]{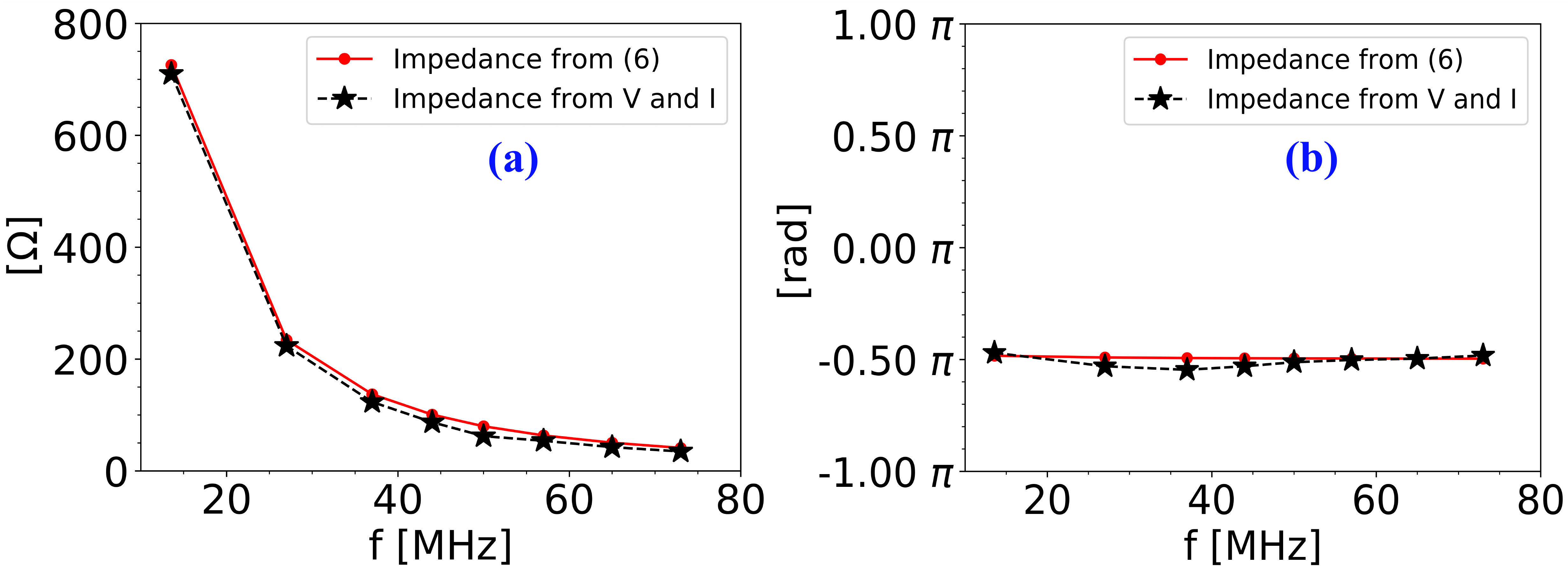}
    \caption{Comparison of the left-hand-sides, calculated using (\ref{eq7:imped}), and the right-hand-sides, obtained from PIC, of equation~(\ref{eq8:Zcomp}) for various driving frequencies. (a) Magnitude, (b) Phase.}
    \label{fig10:Zcomp}
\end{figure*}

Although the simple equivalent circuit model cannot describe the complex phenomena studied in this work, it would nevertheless be useful to infer the equivalent circuit parameters from the detailed PIC simulations so that the discharge could be easily integrated into the overall power circuit. The parameters in equation~(\ref{eq6:circ}) can be calculated if we know the combined sheath thickness and a representative plasma electron density. In the homogeneous model underpinning the equivalent circuit, the electron density profile is assumed to be a step function, with zero value in the sheath and constant value in the quasineutral plasma. It is therefore non-trivial to convert the smooth profile of electron density illustrated in figure~\ref{fig8:denpro} into a step function profile of the homogeneous model. Using a simple approach to extract the homogeneous model parameters from the PIC results, we assume that the combined sheath thickness $d_{\rm sh}$ is the maximum value of the location where the electron density is 10 \% of the ion density over one RF cycle; and that the homogeneous plasma density $n_{\rm e}$ is equal to the PIC-calculated electron density at the center of the discharge gap. The magnitude and phase of the impedance can be compared to the applied voltage and the discharge current obtained from the simulation:
\begin{eqnarray}
|Z| = \frac{V_{\rm 0}}{I_{\rm 0}} \quad {\rm and} \quad \phi_{\rm Z} = {\rm phase\; shift\; between\; I\; and\; V}.
\label{eq8:Zcomp}
\end{eqnarray}
where $V_{\rm 0}$ and $I_{\rm 0}$ are the amplitudes of the voltage and current respectively, and the phase shift between them is obtained by taking the average of the shifts between the maximum peaks and the minimum peaks of the waveforms, since the current is not a purely sinusoidal function. 

Figure~\ref{fig10:Zcomp} shows the comparison of the magnitude and phase of the impedance obtained with equation~(\ref{eq7:imped}) and equation~(\ref{eq8:Zcomp}) for all the values of driving frequencies. The magnitudes match very well for all frequencies, while the phases have a maximum deviation of about 7\%. A sensitivity analysis was conducted using various ratios of electron to ion densities ranging from 1\% to 30\% and the corresponding sheath edge locations to determine the best fit for the left and right-hand-sides of equation~(\ref{eq8:Zcomp}), and it was found that defining the sheath edge with $n_{\rm e}/n_{\rm i} = 10\%$ gives the least deviation. Simple estimates for the combined sheath thickness and the electron density thus give accurate representation of the RF discharge as an electrical circuit. 

\begin{figure}[htb]
    \centering
    \includegraphics[width=0.45\textwidth]{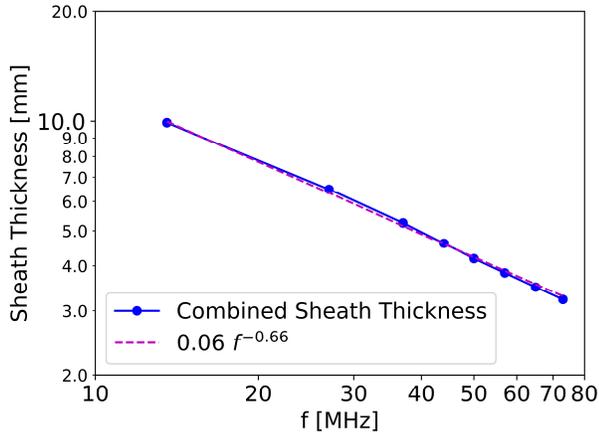}
    \caption{Variation of sheath thickness with frequency and its comparison to a power law fit.}
    \label{fig11:shthick}
\end{figure}

The scaling of sheath thickness with respect to the driving frequency depends on the operating conditions, like driving source (voltage/current/power driven), and the electrode gap. Using a simple model comparing the ion drift flux to the ambipolar diffusion flux, Raizer et al.\cite{raizer1995rfcd} predicted the dependence of sheath thickness on driving frequency as $d_{\rm sh} \propto f^{-0.67}$. Vahedi et. al.\cite{vahedi1993verification} used an electron power balance to predict that the sheath thickness scales approximately inversely with the frequency and confirmed this using two-dimensional PIC simulations for an argon discharge driven by a constant voltage at 60 mTorr. Experimental studies have been done at various driving frequencies on relatively high pressure (around 1~Torr) air discharges driven by a constant power,\cite{khomenko2020capacitively} and by a constant voltage amplitude\cite{khomenko2019capacitively}, and have found the dependence of the sheath thickness calculated using the lumped-parameter model to be $d_{\rm sh} \propto f^{-0.43}$ when driven by constant power, and $d_{\rm sh} \propto f^{-0.67}$ when driven by constant voltage. In our case of a voltage driven discharge, when the estimated combined sheath thickness is plotted against driving frequency (figure~\ref{fig11:shthick}), we see its dependence as $d_{\rm sh} \propto f^{-0.66}$, which is very close to the estimates of $d_{\rm sh} \propto f^{-0.67}$ predicted by the theory.\cite{raizer1995rfcd}

\begin{figure}[htb]
    \centering
    \includegraphics[width=0.45\textwidth]{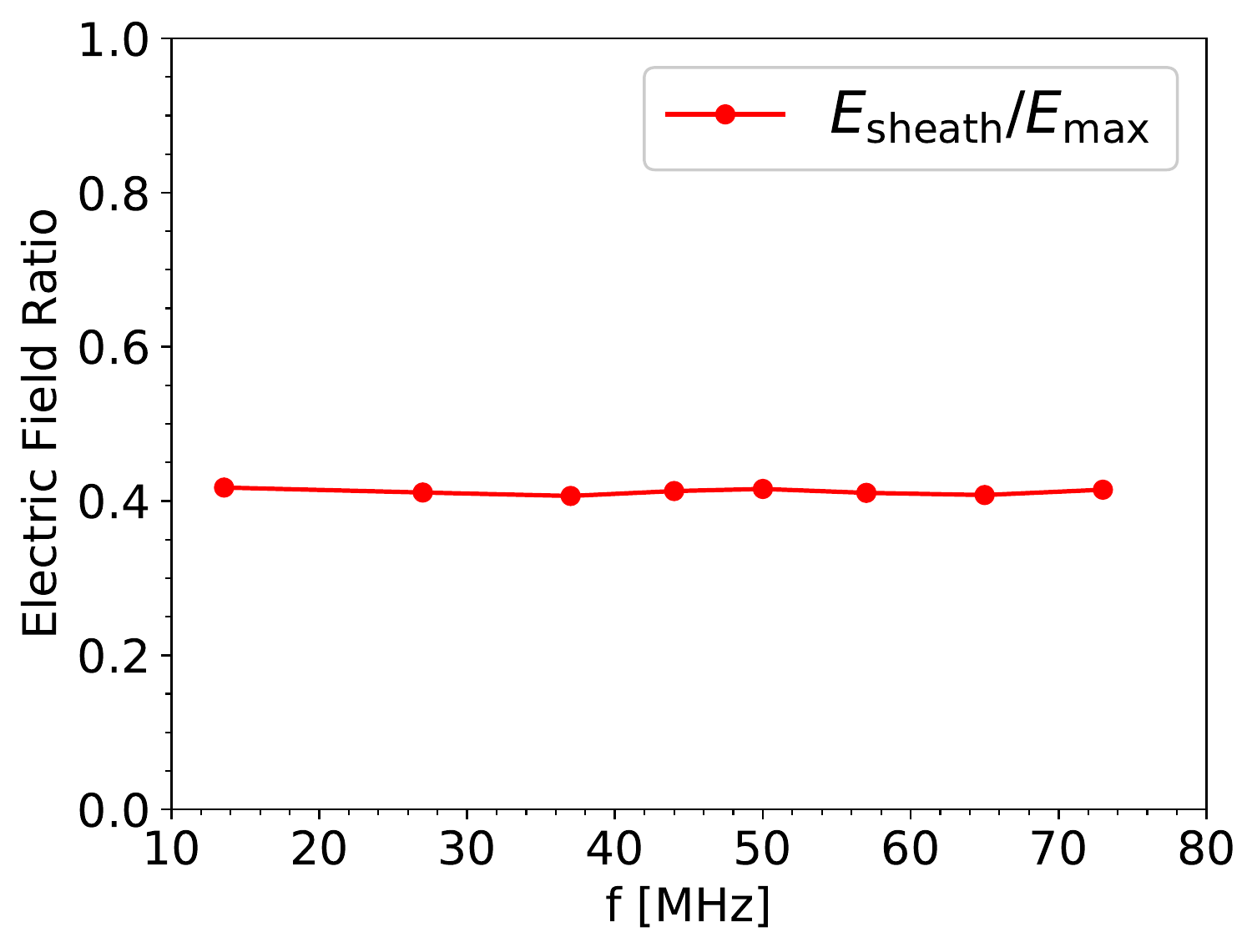}
    \caption{Ratio of electric field at the sheath edge to that at the powered electrode at various driving frequencies.}
    \label{fig12:Eratio}
\end{figure}

Defining the sheath edge at the location where the electron density is 10\% of the ion density gives a good estimate of the thickness of the sheath for the equivalent circuit model. Another way to define the sheath thickness is comparing the electric field at the sheath edge to the electric field at the electrode. In our study, we found that at the value of sheath thickness used for the calculations in equation~(\ref{eq6:circ}), the ratio of the electric field at that location to the field at the electrode remains constant (about 0.4) irrespective of the driving frequency, as shown in figure~\ref{fig12:Eratio}.

While the lumped-parameter model gives a simple way to estimate sheath thickness, it does not take into account more complex phenomena happening at the sheath edge. The spatial variation of ion density and the production of electron beams in the sheath modulate the sheath edge such that it oscillates with higher harmonics in addition to the driving frequency.\cite{sharma2016effect,sharma2019electric,wen2019secondary,rauf2009power,schulze2011ionization} Investigating these nonlinear effects will aid in understanding the discharge dynamics.

\subsection{\label{sec:j_e.E}Electric Field and Electron Heating}

The power deposition to the electrons was computed by time averaging $\boldsymbol{E} \! \cdot \! \boldsymbol{j}_{\rm e}$ over the last 100 RF cycles after the simulations reached a steady state. Deposited power is transferred to the kinetic energy of electron motion, which in turn is randomized in part in collisions, leading to increased electron temperature. Negative values of the power deposition indicate that the electrons lose kinetic energy to the field.

\begin{figure*}[htb]
    \centering
    \includegraphics[width=0.9\textwidth]{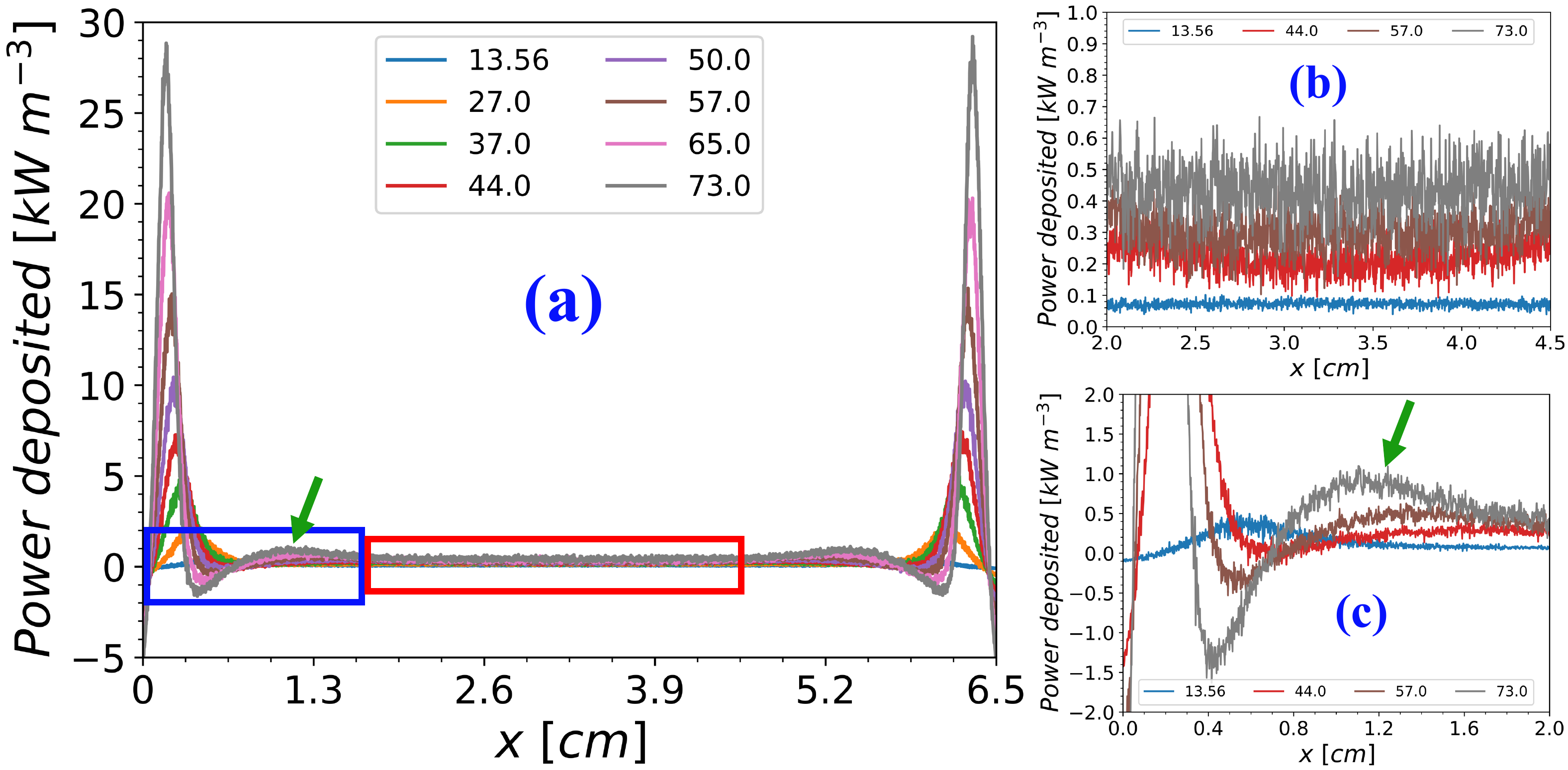}

    \caption{Time averaged electron heating $\left< \boldsymbol{E} \! \cdot \! \boldsymbol{j}_{\rm e} \right>$ for (a) the entire domain, with some selected zoomed in profiles denoted by (b) the red box at the center, and (c) the blue box in the vicinity of the sheath edge, for a range of driving frequencies. The green arrows show the local maxima of power deposition near the sheath edge.}
    \label{fig13:jE}
\end{figure*}

Figure~\ref{fig13:jE} shows the profiles of average power deposition $\left< \boldsymbol{E} \! \cdot \! \boldsymbol{j}_{\rm e} \right>$ for different driving frequencies. Electrons get heated in vicinity of the sheath; this energy deposition increases as the frequency increases as shown in figure~\ref{fig13:jE}(a). Since electron density in the sheath is negligible compared to that in the bulk plasma, the electrons gain energy on an average in a collisionless manner  through interaction with the oscillating, highly nonlinear electric field rather than through ohmic heating. In such a case, the energy gained depends on the velocity of oscillation of the plasma-sheath boundary. At high driving frequencies, the plasma-sheath boundary oscillates much faster compared to the low driving frequency cases. Hence, we have higher power deposition at high driving frequencies. Later in this section, we will see that there are higher harmonic oscillations present in the sheath, in addition to the fundamental driving frequency. These nonlinearities enhance power deposition by the field into the electrons for the high driving frequency cases.

\begin{figure*}
    \centering
    \includegraphics[width=0.8\textwidth]{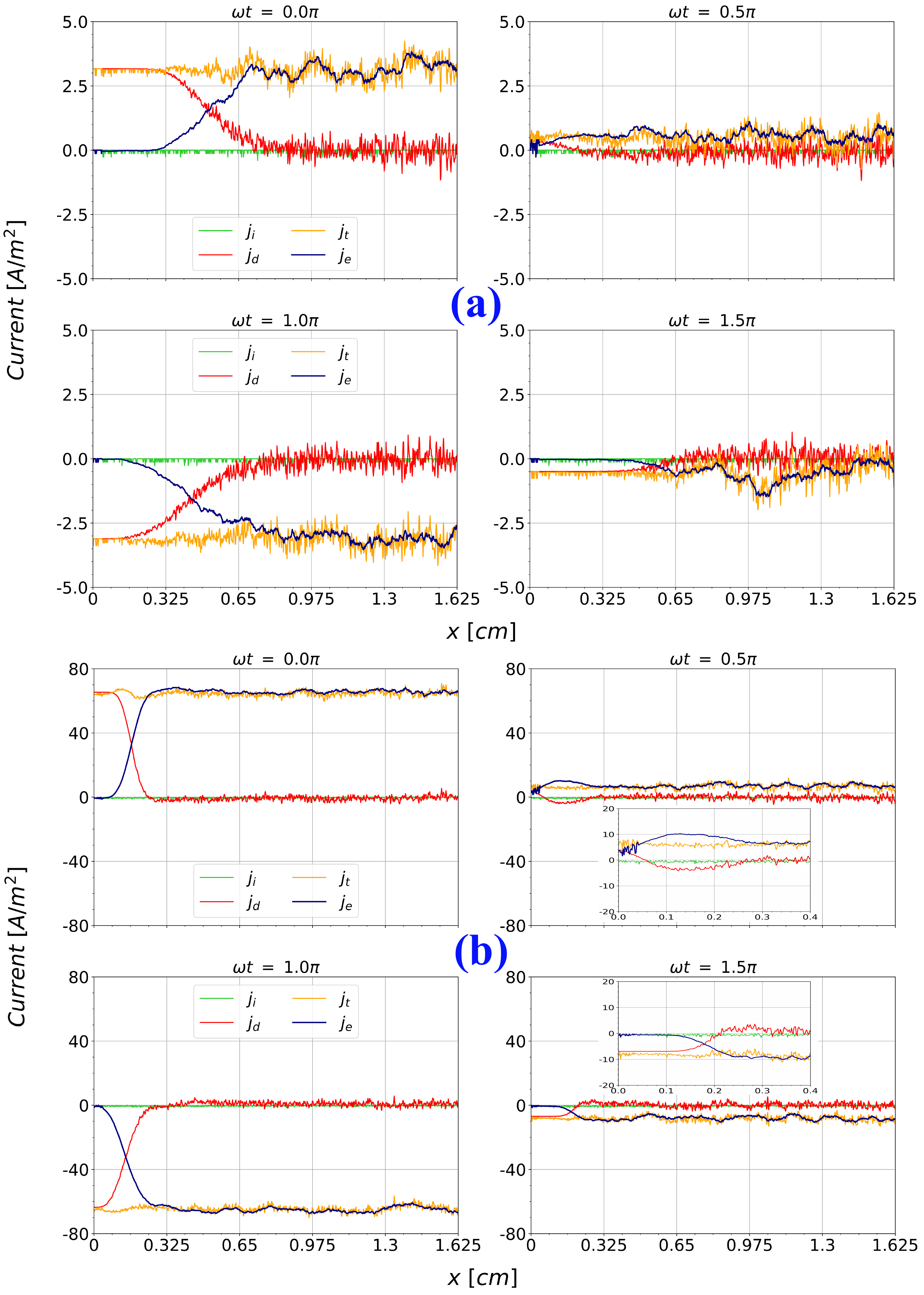}
    \caption{Electron, ion, displacement, and total current near the sheath, conditionally averaged on phase, at (a) 13.56~MHz and (b) 73~MHz.}
    \label{fig14:curr}
\end{figure*}
As the driving frequency increases, the amplitude of electron oscillation decreases. Hence, at higher driving frequencies, the electrons lose their energy in a narrower region near the sheath edge, compared to the lower driving frequency cases, as seen in figure~\ref{fig13:jE}(a). In the center of the discharge, $\left< \boldsymbol{E} \! \cdot \! \boldsymbol{j}_{\rm e} \right>$ is relatively small due to the lower values of electric field there. The power deposition in the center of the discharge gap increases as frequency increases, as seen in figure~\ref{fig13:jE}(b), where selected cases from the region denoted by the red box in figure~\ref{fig13:jE}(a) are plotted. This is due to the increase in center electron density as the frequency increases, giving rise to a larger electron current.

Careful consideration is required to assess the power deposition by the field to the electrons near the sheath edge. As we observe in figure~\ref{fig13:jE}(c), where selected cases from the region denoted by the blue box in figure~\ref{fig13:jE}(a) are plotted, just outside the sheath edge electrons lose energy on average to the field, an effect which increases in magnitude as the frequency increases. Closer to the center from this region of negative $\left< \boldsymbol{E} \! \cdot \! \boldsymbol{j}_{\rm e} \right>$, there is a local maximum in power deposition for the high driving frequency data shown by the green arrows in figure~\ref{fig13:jE}. 

CCP discharges operated by both sinusoidal and non-sinusoidal waveforms have shown to produce fast electron beams during the sheath expansion phase.\cite{wilczek2015effect,wilczek2018disparity,sharma2020electric,sharma2020driving,sharma2018spatial,sharma2020high,sharma2022plasma} At low gas pressures, these electron beams travel uninhibited to the opposite sheath. If the opposite sheath is in the collapsed phase when the fast electron beams reach it, the electrons are lost to the opposite electrode and are not confined in the plasma, as was shown by Wilczek et. al.\cite{wilczek2015effect} On the other hand, if the electron beams reach the opposite sheath during its expanded phase, they are reflected back into the plasma and cause additional ionization in the bulk. Researchers have observed that at low pressures there is a critical driving frequency above which the electron beams are successfully confined in the plasma and cause a step increase in the electron density at the center of the plasma.

As mentioned before, Wilczek et. al.\cite{wilczek2015effect} also studied the effect of background gas pressure on the dependence of bulk electron density versus the driving frequency and found that at pressures above 2~Pa (about 15~mTorr) the drastic increase in electron density disappeared, and the trend followed was closer to the theoretical approximation of $n_{\rm e} \sim f^2$, for which non-local kinetic effects are ignored.\cite{lieberman2005principles,chabert_braithwaite_2011}

Figures~\ref{fig14:curr}(a) and (b) show the different components of current for 13.56 MHz and 73 MHz respectively. Here, the currents are conditionally averaged based on the phase of the voltage input. They are plotted at four different phases of an RF cycle, specifically $\omega t = 0$, $0.5 \pi$, $1.0 \pi$, and $1.5 \pi$, based on the sinusoidal voltage waveform defined in section~\ref{sec2:procedure}.

At 50~mTorr gas pressure, we see the effect of fast electron beams near the sheath edge as a local increase in electron conduction current ($j_{\rm e}$), about 71\% greater that the total current ($j_{\rm t}$) value, as shown in figure~\ref{fig14:curr}(b) at $\omega t = 0.5 \pi$ and $1.5 \pi$. Since the total current (conduction plus displacement) should remain constant over the entire domain, the increase in $j_{\rm e}$ is compensated by an equivalent change in displacement current, $j_{\rm d}$. We see from figure~\ref{fig14:curr}(a) that the local increase in $j_{\rm e}$ is absent at low driving frequencies and is only observed as the frequency increases. Since the higher gas pressure hinders uninhibited travel of the electron beams, they are unable to reach the opposite sheaths. Instead, local heating is observed up to short distance after the sheath edge, about 8~mm to 16~mm as seen in figure~\ref{fig13:jE}(c), before the electrons lose their energy in inelastic collisions.
\begin{figure*}[htb]
    \centering
    \includegraphics[width=0.9\textwidth]{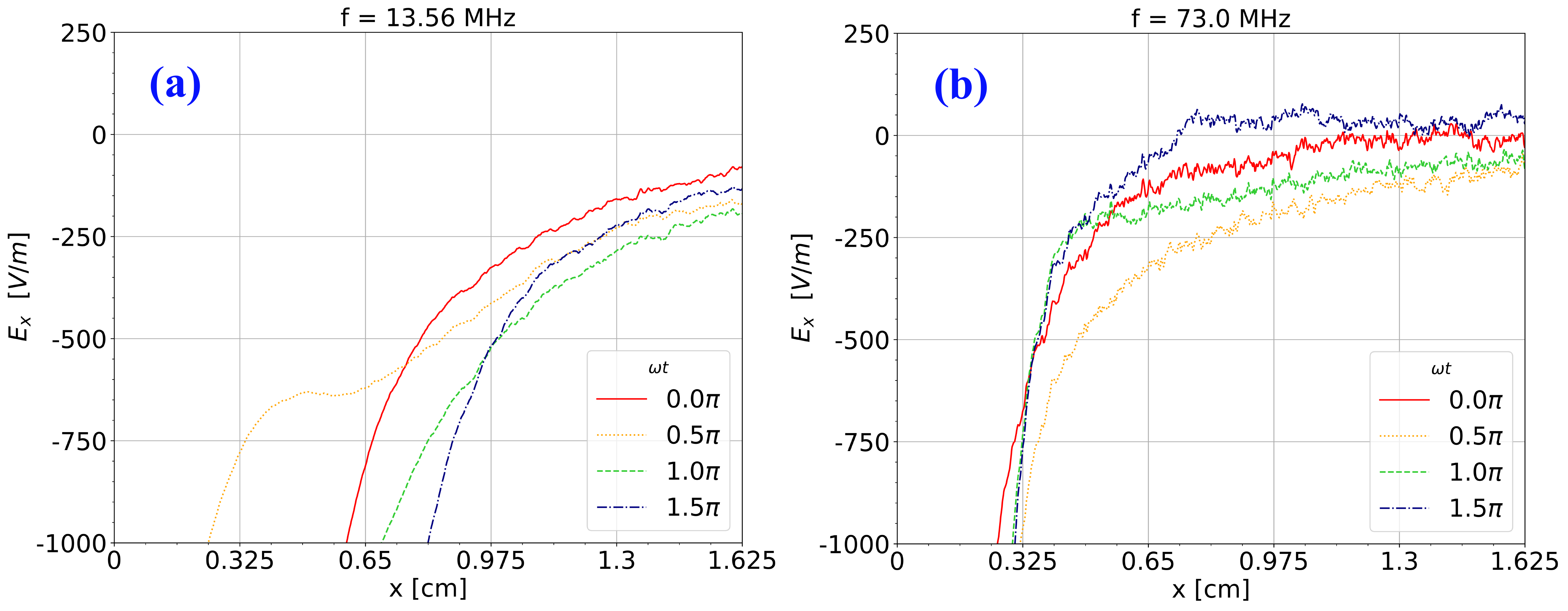}
    \caption{Electric field, conditionally averaged on phase, near the powered electrode over one RF cycle at (a) 13.56~MHz and (b) 73~MHz.}
    \label{fig15:field}
\end{figure*}

As the fast electron beams exiting the sheath travel through the plasma, colliding with the neutral gas, they encounter a region of opposing electric field just after the sheath edge. Figure~\ref{fig15:field} shows the conditionally averaged (based on the phase of the voltage input) electric field near the left (powered) electrode at the same phases in the RF cycle as the plots of current in figure~\ref{fig14:curr}. At $\omega t = 0.5 \pi$, the sheath is in the collapsed phase, top right plot of figure~\ref{fig14:curr}(b), and the increase in local conduction current is positive, which results from the fast electrons moving towards the electrode. As seen in figure~\ref{fig15:field}(b), the electric field at this moment in the same region is negative, which results in an electric force on the electrons in the direction away from the electrode. Similarly, at $\omega t = 1.5 \pi$, the sheath is in the expanded phase, bottom right plot of figure~\ref{fig14:curr}(b), and the increase in local conduction current is negative, which is again opposite to the electric force on the electrons, with positive field in figure~\ref{fig15:field}(b). Averaging this over the entire RF cycle results in a negative $\left< \boldsymbol{E} \! \cdot \! \boldsymbol{j}_{\rm e} \right>$ just outside the sheath edge. At 13.56 MHz driving frequency, we do not observe the production of fast electron beams near the sheath edge, since there is no local increase in the electron conduction current as seen in figure~\ref{fig14:curr}(a).

\begin{figure*}[htb]
    \centering
    \includegraphics[width=0.9\textwidth]{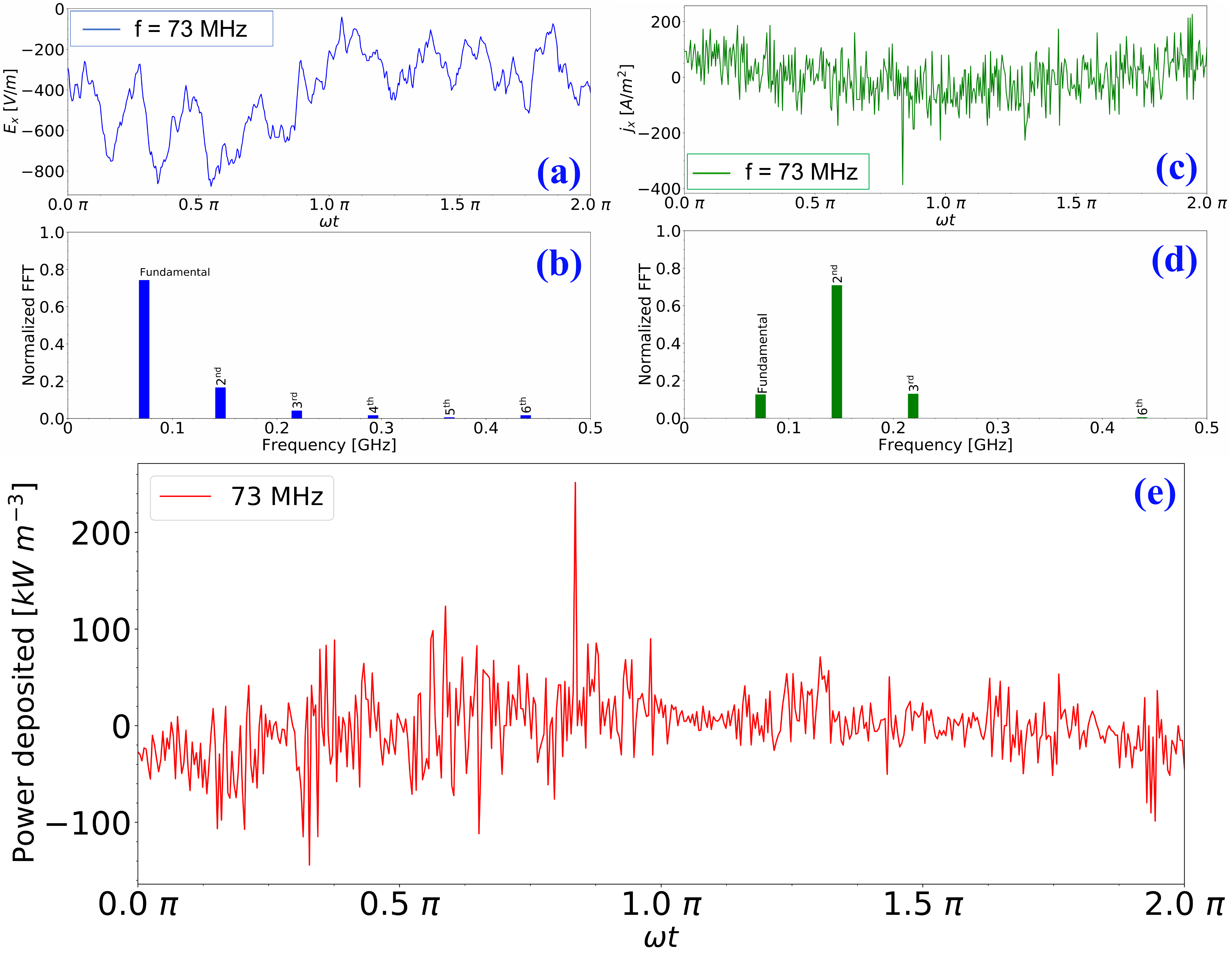}
    \caption{Properties at the location of minimum $\left< \boldsymbol{E} \! \cdot \! \boldsymbol{j}_{\rm e} \right>$ for the 73~MHz driving frequency case. (a) Time variation of the electric field over one RF cycle; (b) Normalized FFT of the electric field; (c) Time variation of the electron current over one RF cycle; (d) Normalized FFT of the electron current; (e) Time variation of $\boldsymbol{E} \! \cdot \! \boldsymbol{j}_{\rm e}$ over one RF cycle}
    \label{fig16:ExFFTf73}
\end{figure*}

\begin{figure*}[htb]
    \centering
    \includegraphics[width=0.9\textwidth]{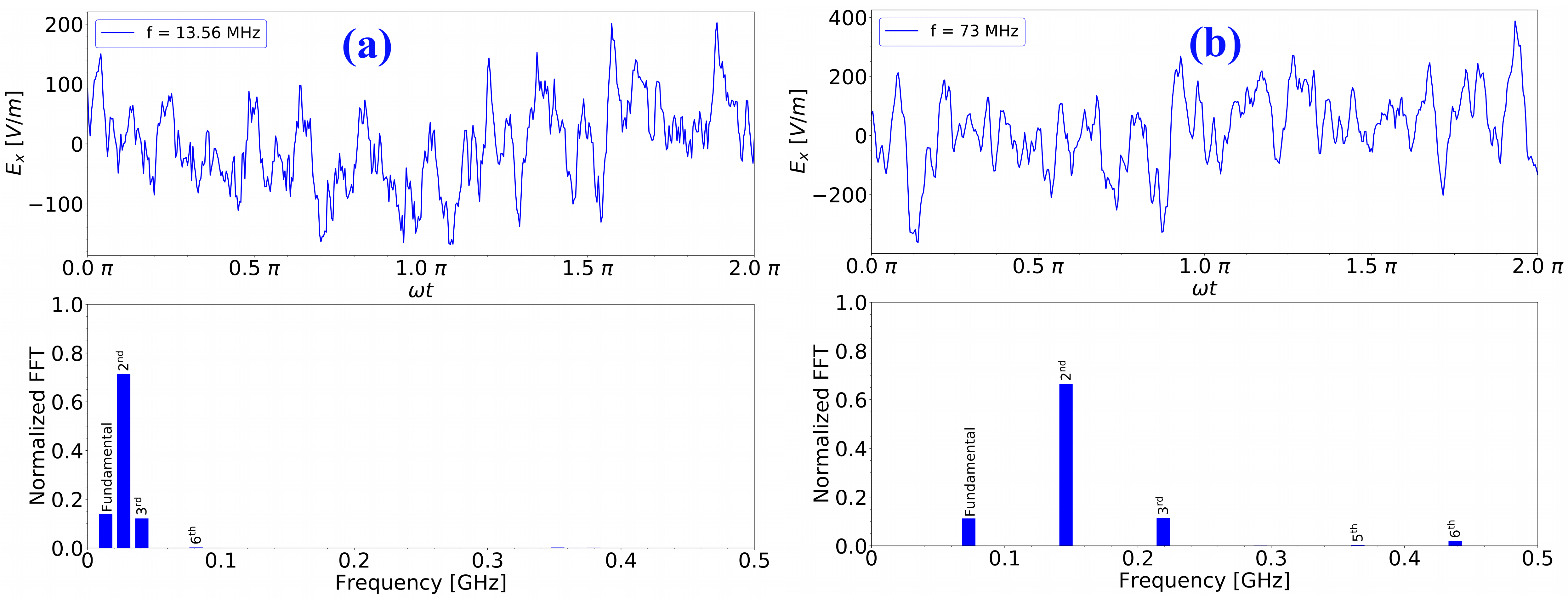}
    \caption{Electric field at the center of the discharge and the corresponding energy normalized FFT for the (a)13.56 MHz and (b) 73~MHz driving frequency case.}
    \label{fig17:ExFFTcenter}
\end{figure*}

Nonlinearities play an important role in the generation of fast electrons, electron cooling near the sheath edge, the eventual local heating after the sheath edge, and the broad positive power deposition plateau around the center of the discharge gap. Nonlinear dynamics in the discharge show up in the form of various high frequency oscillations of properties like the electric field and electron current. Figures~\ref{fig16:ExFFTf73}(a) and (b) show the time variation of the electric field in one RF cycle of the 73~MHz case at the location of minimum $\left< \boldsymbol{E} \! \cdot \! \boldsymbol{j}_{\rm e} \right>$ (around 4~mm) and the electric field at that location in the frequency domain. Figures~\ref{fig16:ExFFTf73}(c) and (d) show the same for the electron current and figure~\ref{fig16:ExFFTf73}(e) shows the variation of $\boldsymbol{E} \! \cdot \! \boldsymbol{j}_{\rm e}$ over one RF cycle at the same location. The vertical axis in figures~\ref{fig16:ExFFTf73}(b) and (d) are normalized such that the energy of all the oscillating components add up to one. The phase difference between the electric field and the electron current is clearly visible in figures~\ref{fig16:ExFFTf73}(a) and (c). From figure~\ref{fig16:ExFFTf73}(b) and (d), we see that along with oscillation at the fundamental frequency (74.2 \% of total wave energy), the second (16.6 \%), third (4.2 \%), fourth (1.7 \%), and the sixth (1.7 \%) harmonics are also present in the electric field, while the electron current oscillates at the second (70.9 \% of total wave energy), third (13 \%), and the sixth (0.4 \%) harmonics, along with the fundamental frequency (12.6 \%). 

The penetration of the higher harmonic oscillation into the quasineutral plasma depends on the local electron plasma frequency, the collision frequency, and the frequency of oscillation of the field. A complex relative permittivity can be calculated for the plasma as:\cite{lieberman2005principles}

\begin{equation}
    \epsilon_{\rm r} = 1 - \frac{\omega_{\rm p}^2}{\omega^2 + \nu_{\rm m}^2} - i \frac{\nu_{\rm m}}{\omega} \frac{\omega_{\rm p}^2}{\omega^2 + \nu_{\rm m}^2}.
    \label{eq9:perm}
\end{equation}

If the real part of equation~(\ref{eq9:perm}) is greater than the imaginary part, then the field wave propagates into the bulk plasma as an evanescent wave whose amplitude reduces with distance. The characteristic penetration depth, $\delta$, can be calculated as: 

\begin{equation}
    \delta = \frac{c}{\omega \sqrt{|\Re\left(\epsilon_{\rm r}\right)|}}. \label{eq10:delta}
\end{equation}

\begin{table}
\caption{\label{tab2:skin}Calculated characteristic penetration depth for oscillating electric field for the 73 MHz case.}
\footnotesize
\begin{ruledtabular}

\begin{tabular}{ll}
Frequency of oscillation (MHz)   &   Characteristic Penetration Depth (cm)\\
\hline
73 (Fundamental)                                    &   7.83\\
146 (2\textsuperscript{nd} harmonic)                &   7.94\\
219 (3\textsuperscript{rd} harmonic)                &   8.25\\
292 (4\textsuperscript{th} harmonic)                &   8.75\\
438 (6\textsuperscript{th} harmonic)                &   10.91\\

\end{tabular}
\end{ruledtabular}
\end{table}
\begin{figure*}[htb]
    \centering
    \includegraphics[width=0.9\textwidth]{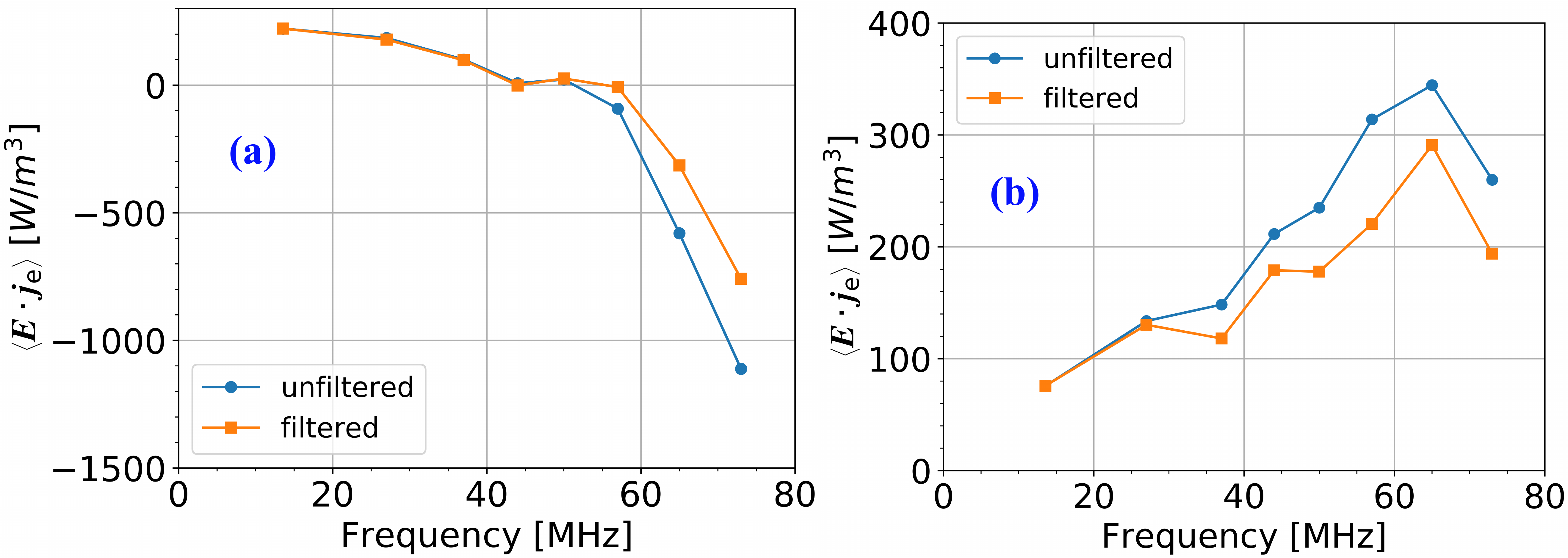}
    \caption{Effect of low-pass filter on computed $\left< \boldsymbol{E} \! \cdot \! \boldsymbol{j}_{\rm e} \right>$ at (a) the location of its minimum, (b) the center of the discharge gap.}
    \label{fig18:jE_fuf}
\end{figure*}
\noindent where c is the speed of light, and $\Re\left(\epsilon_{\rm r}\right)$ denotes the real part of equation~(\ref{eq9:perm}). For the 73 MHz driving frequency case, the calculated $\Re\left(\epsilon_{\rm r}\right)$ dominates for the frequencies observed in figure~\ref{fig16:ExFFTf73} at the local electron plasma frequency. The calculated characteristic penetration depths for the fundamental frequency and the higher harmonics involved are shown in table~\ref{tab2:skin}. The values of $\delta$ at all the frequencies are greater than the discharge gap, justifying the of a quasi-electrostatic model in the present work. These oscillating fields thus penetrate the entire plasma domain and demonstrate the nonlinearity of the discharge as a whole. 

In general, as we move from the sheath edge to the center of the plasma, these nonlinearties interact to either enhance or reduce the different components of oscillation. Figures~\ref{fig17:ExFFTcenter}(a) and (b) show that that at both low and high driving frequencies, the 2\textsuperscript{nd} harmonic oscillation of the electric field is much more significant than the fundamental frequency at the center of the discharge. We have only observed oscillation in the electric field at a few harmonics higher than the fundamental frequency for our 50 mTorr case. The high number of collisions in the discharge gap suppresses very high frequency oscillation. Research has demonstrated that low pressure discharges (less than 10 mTorr) display harmonics at very high frequencies (larger than 10 or 20 times the fundamental frequency) at the center of the gap.\cite{sharma2019influence,sharma2020driving,sharma2019electric,wilczek2018disparity}

As an example of the role these harmonics play at higher driving frequencies, the electric field and the current were low-pass filtered so that only the fundamental frequency was retained and $\left< \boldsymbol{E} \! \cdot \! \boldsymbol{j}_{\rm e} \right>$ was recalculated from the filtered values. Figure~\ref{fig18:jE_fuf} shows the comparison of $\left< \boldsymbol{E} \! \cdot \! \boldsymbol{j}_{\rm e} \right>$ at the location of minimum $\left< \boldsymbol{E} \! \cdot \! \boldsymbol{j}_{\rm e} \right>$ and at the center of the discharge gap using unfiltered and filtered field and current. We see from figure~\ref{fig18:jE_fuf}(a) that the electrons lose energy only when the driving frequency is 50~MHz and higher. At frequencies below 50~MHz, filtering the current and the field does not have much effect on $\left< \boldsymbol{E} \! \cdot \! \boldsymbol{j}_{\rm e} \right>$, whereas above 50~MHz, including oscillations only at the fundamental frequency drastically reduces the magnitude of electron cooling.

Nonlinearities contribute to local electron heating in the center of the discharge gap as well. From figure~\ref{fig18:jE_fuf}(b), we see that at higher driving frequencies, the magnitude of electron heating at the center is significantly lower when the filtered values of current and electric field were used.

The presence of electric field reversal near the sheath edge and its effect on power deposition in low pressure collisionless discharges have been reported in literature.\cite{sato1990electron,turner1992anomalous,czarnetzki1999space,gans2003phase,sharma2013simulationrever,sharma2013simulation,schulze2008electric,o2008plasma,meige2008plasma,sharma2013critical,sharma2014investigation} Schulze et. al.\cite{schulze2008electric} investigated electric field reversals near the sheath edge in single-frequency discharges and compared their simulation results to a simple fluid model they developed for the electric field. They used the momentum equation for electrons in the discharge, along with the electron mass conservation equation, to obtain an expression for the electric field as:
\begin{equation}
    E = \frac{m}{ne^{2}} \left(
        \overbrace{\frac{\partial j_{\rm e}}{\partial t}}^{\text{1}} + \overbrace{\nu_{\rm c} j_{\rm e}}^{\text{2}} + 
        \overbrace{j_{\rm e}^{2} \frac{1}{en^{2}} \frac{\partial n}{\partial x}}^{\text{3}} - 
        \overbrace{j_{\rm th}^{2} \frac{1}{en^{2}} \frac{\partial n}{\partial x}}^{\text{4}}
        \right). \label{eq11:E_rev}
\end{equation}
Here, $m$ and $e$ are the mass and charge of the electron respectively, $\nu_{\rm c}$ is the collision frequency, $n$ is the local electron density, $j_{\rm e}$ is the electron current, and $j_{\rm th}^{2} = e^{2}n^{2}kT_{\rm e}/m$ is a current-like term due to diffusion of electrons.

The first term on the right-hand-side of equation~(\ref{eq11:E_rev}) corresponds to the field due to temporal changes in the electron current and is attributed to electron inertia. The third term is also an inertial term (the convective acceleration) which is attributed to the spatial changes in electron density. The second term accounts for collisions and the fourth term corresponds to diffusion.

\begin{figure*}[htb]
    \centering
    \includegraphics[width=0.9\textwidth]{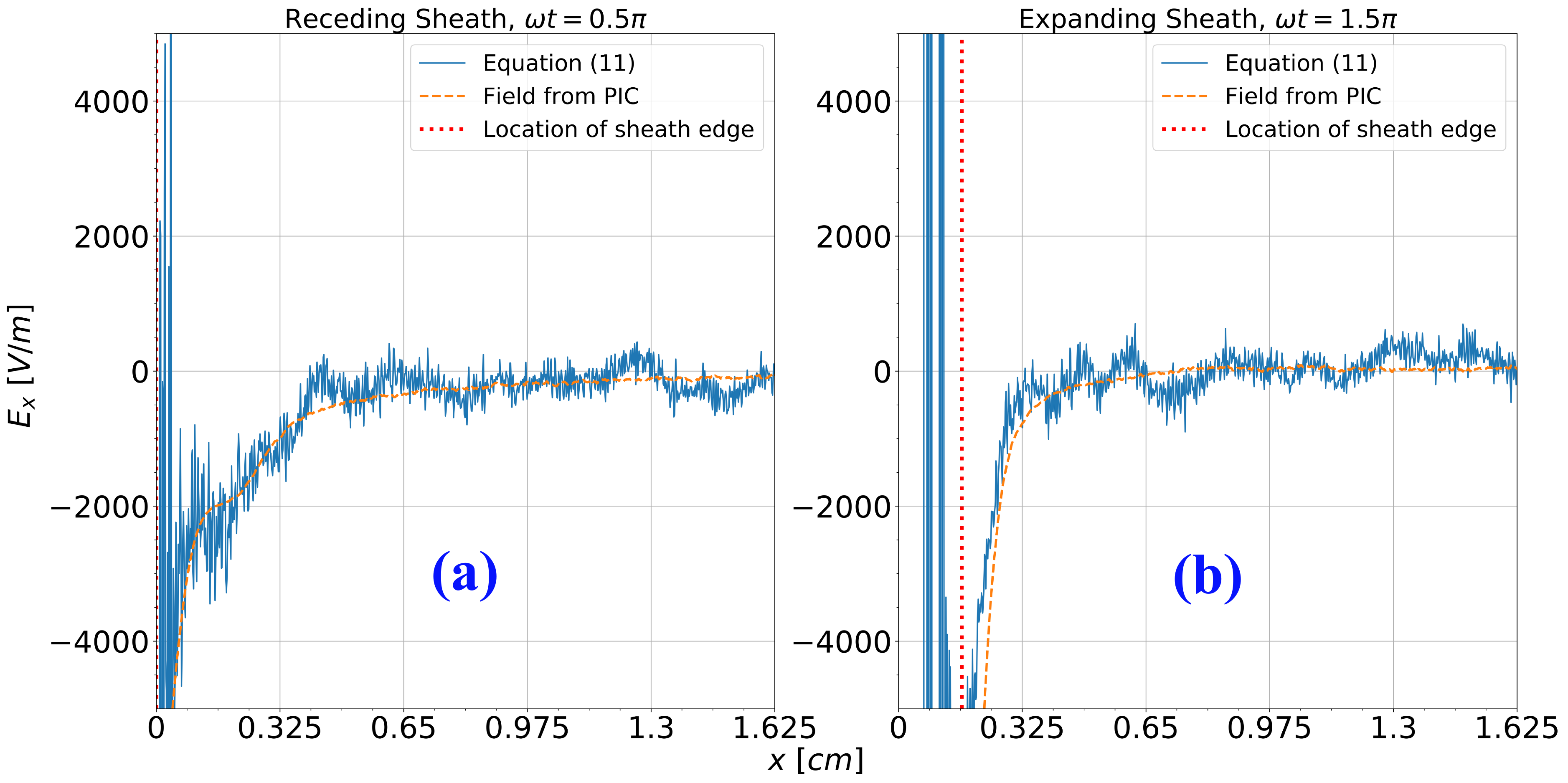}
    \caption{Comparison of electric field profiles obtained directly from PIC (73~MHz driving frequency) to equation~(\ref{eq11:E_rev}) for (a) receding sheath and (b) expanding sheath.}
    \label{fig19:Field_rev}
\end{figure*}

This is a simplified expression obtained under a number of assumptions. It is valid only in the quasineutral region of the plasma where electron conduction current is dominant over the displacement current. It does not account for the oscillation of the sheath. The electron density profile is assumed to be a step function which increases from zero in the sheath to the plasma density at the sheath edge. Ionization in the sheath is neglected by assuming sheath width much smaller than the bulk width, which eliminates the source term in the equation for the conservation of mass of electrons. When this simple analytical expression is plotted using current and electron density from the simulation and compared to the electric field obtained directly from PIC (figure~\ref{fig19:Field_rev}), they match very well in the bulk of the plasma and near the sheath edge, but both the curves do not show signs of field reversal as was found in collisionless discharges. This was true for all the driving frequencies studied.

\begin{figure*}[htb]
    \centering
    \includegraphics[width=0.9\textwidth]{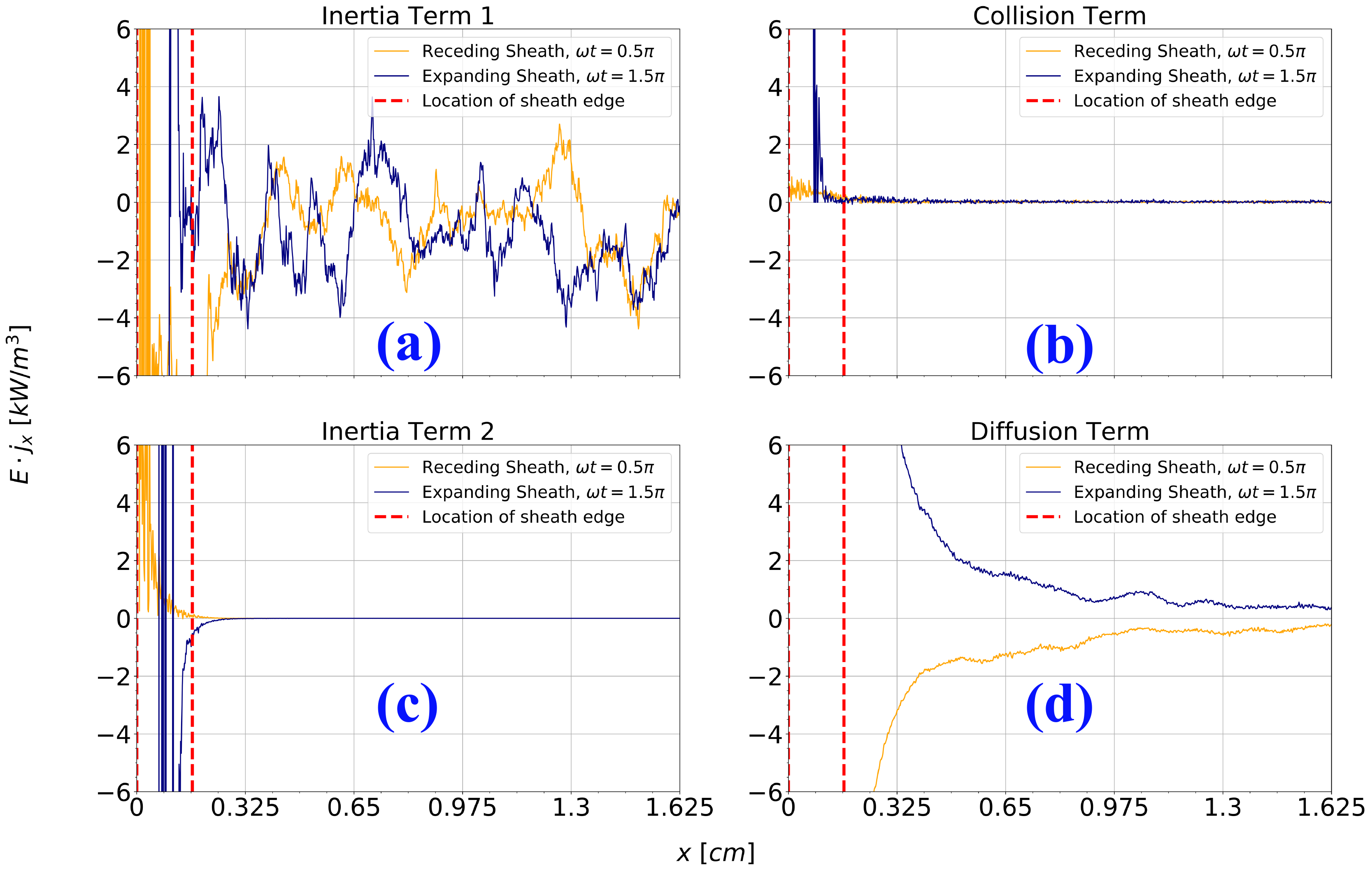}
    \caption{Contribution of (a) inertia due to temporal changes in current, (b) collisions, (c) inertia due to spatial changes in electron density, and (d) diffusion of electrons in equation~(\ref{eq11:E_rev}), to the electric field for 73~MHz driving frequency.}
    \label{fig20:Field_cont}
\end{figure*}

Using equation~(\ref{eq11:E_rev}) as the representation of the electric field in the discharge, we can analyze the contribution of the individual terms in the equation to the average power deposition. Figure~\ref{fig20:Field_cont} shows zoomed in profiles of the individual terms of equation~(\ref{eq11:E_rev}) multiplied by the electron current which correspond to power deposition (a) from electron inertia due to instantaneous changes in the electron current (which we will call inertia~term~1), (b) owing to the effect of collisions (collision~term), (c) from inertia due to changes in the electron density profile (inertia~term~2), and (d) owing to the effect of diffusion of electrons (diffusion~term). The red and orange curves correspond to the same phase in the RF cycle as the red and orange curves in figure~\ref{fig15:field}(b).

Past the sheath edge, the contributions of inertia~term~2 and the collision~term are negligible. During both halves of the cycle, irrespective of the potential at the powered electrode, inertia~term~1 contributes negative values to the average power deposition, as seen in figure~\ref{fig20:Field_cont}(a). In a perfectly sinusoidal system, averaging the term $\frac{\partial \boldsymbol{j}}{\partial t} \cdot \boldsymbol{j}$ over an RF cycle should equate to zero. In our case of a highly nonlinear system, the contribution from inertia~term~1 to the average power deposition remains negative. While the diffusion~term also significantly contributes to $\boldsymbol{E} \cdot \boldsymbol{j}_{\rm e}$ as seen in figure~\ref{fig20:Field_cont}(d), it oscillates between positive and negative values during different phases of the cycle. Therefore, electron inertia, as a result of fast electron beams disrupting the linearity of the discharge, is responsible for electron cooling near the sheath edge.

Thus, simulations and analysis show that the negative heating of electrons near the sheath edge at the conditions considered here is not associated with field reversal. Instead, it is the result of electron inertia in rapidly oscillating fields combined with the discharge nonlinearity.

\section{Summary and Conclusions}

A computational study using PIC methods was carried out of CCP discharges in argon at 50~mTorr and driving frequencies from 13.56~MHz to 73~MHz. The energy distributions at the center of the plasma were complicated and non-Maxwellian for all driving frequencies. When compared to published experimental results, the distributions matched reasonably well in a mid-energy range, 2~eV to 10~eV. Outside that range, the experiment data may be unreliable due to the resolution limitations of the probe used in the experiments. The predictions matched the experimental trend of increasing electron density and decreasing electron temperature at the center of the discharge.

In the context of a lumped-parameter circuit model, the calculated impedance of the discharge matched well with the same parameter calculated through the voltage and current from the simulation. The estimates of the sheath thickness used to obtained the impedance values at all the frequencies followed the power law trend of $d_{\rm sh} \propto f^{-0.66}$. The ratio of electron to ion number densities at the best estimates of the sheath edge location was 10\% and the ratio of the electric field at the estimated sheath edge to the maximum electric field revealed a constant value of 0.4. These two ratios can be used to approximate the sheath thickness for the equivalent circuit model.

Interesting features were observed in profiles of averaged power deposition to the electrons $\left< \boldsymbol{E} \! \cdot \! \boldsymbol{j}_{\rm e} \right>$. Power deposition increases with increasing frequency in the sheath due to enhanced heating of electrons at high driving frequencies. Electron heating at the center of the discharge increases with increasing driving frequency due to the increased conductivity of the plasma at higher frequencies, aided by higher harmonic oscillation of the electric field. Near the sheath edge, power deposition first decreases, becomes negative, and then increases above zero, before finally reducing to the value at the center. The negative $\left< \boldsymbol{E} \! \cdot \! \boldsymbol{j}_{\rm e} \right>$ past the sheath edge is due to the production of fast electron beams in the sheath which travel in a direction against the electrical force on the electrons giving rise to electron cooling there. After this region of cooling, the electron beams aid in enhanced heating before losing their energy in inelastic collisions. The production of fast electron beams is due to the nonlinear dynamics of the sheath. The fast oscillations in the sheath, and consequently the field, energize the electrons on a timescale much smaller than the time required for the electrons to react, adding to the electron heating in the sheath, increasing electron cooling near the sheath edge, and enhancing the electron heating in the bulk of the plasma.

A simple theoretical approach derived from the momentum equation was used to analyze the contributions from various plasma phenomena to the behavior of electron heating. The analysis showed that electron inertia plays a significant role, resulting, in combination with harmonic generation, in an overall average negative power deposition near the sheath edge.

\begin{acknowledgments}
    This work was conducted as a collaborative research project at the Princeton Collaborative Research Facility (PCRF), which is supported by the U.S. Department of Energy (DOE) under Contract No. DE AC02-09CH11466. The work by Purdue University co-authors (S.S., J.P., and S.M.) was supported by the US DOE Fusion Energy Sciences Office under the grant DE-SC0021076 (Dr. Nirmol Podder, Program Manager). We express our gratitude to Purdue University's Rosen Center for Advanced Computing (RCAC) for computer resources to run the simulations.
\end{acknowledgments}

\nocite{*}
\bibliography{references}

\end{document}